\documentclass[fleqn,usenatbib]{mnras}

\usepackage{newtxtext,newtxmath}

\usepackage[T1]{fontenc}

\DeclareRobustCommand{\VAN}[3]{#2}
\let\VANthebibliography\thebibliography
\def\thebibliography{\DeclareRobustCommand{\VAN}[3]{##3}\VANthebibliography}

\usepackage{graphicx}	
\usepackage{amsmath}	

\def\aj{AJ}%
\def\araa{ARA\&A}%
\def\apj{ApJ}%
\def\apjl{ApJ}%
\def\aap{A\&A}%
\def\icarus{Icarus}%
\def\mnras{MNRAS}%
\def\solphys{Sol.~Phys.}%
\def\ssr{Space~Sci.~Rev.}%
\def\nat{Nature}%
\def\iaucirc{IAU~Circ.}
\def\jqsrt{J.~Quant.~Spec.~Radiat.~Transf.}%
\def\planss{Planet.~Space~Sci.}%
\def\philtransa{Phil.~Trans.~R.~Soc.~A}%

\title[Polarimetric observations of comet C/2010 E6]{Polarimetric analysis of \textit{STEREO} observations of sungrazing Kreutz comet C/2010 E6 (STEREO)}

\author[R. Ne\v{z}i\v{c} et al.]{
Rok Ne\v{z}i\v{c},$^{1,2,3}$\thanks{E-mail: rok.nezic@armagh.ac.uk (RN)}
Stefano Bagnulo,$^{1}$
Geraint H. Jones,$^{2,3}$
Matthew M. Knight$^{4,5}$  
\newauthor
and Galin Borisov$^{1,6}$
\\
$^{1}$Armagh Observatory and Planetarium, College Hill, Armagh, BT61 9DG, UK \\
$^{2}$Mullard Space Science Laboratory, University College London, Holmbury St. Mary, Dorking RH5 6NT, UK \\
$^{3}$The Centre for Planetary Sciences at UCL/Birkbeck, Gower Street, London WC1E 6BT, UK \\ 
$^{4}$Department of Physics, United States Naval Academy, 572C Holloway Rd, Annapolis, MD 21402, USA \\
$^{5}$Department of Astronomy, University of Maryland, College Park, MD 20742, USA \\
$^{6}$Institute of Astronomy and National Astronomical Observatory, Bulgarian Academy of Sciences, 72, Tsarigradsko Chauss\`{e}e Blvd., Sofia BG-1784, Bulgaria}

\date{Accepted XXX. Received YYY; in original form ZZZ}

\pubyear{2022}

\begin{document}
\label{firstpage}
\pagerange{\pageref{firstpage}--\pageref{lastpage}}
\maketitle
\begin{abstract}
Twin \textit{STEREO} spacecraft pre-perihelion photometric and polarimetric observations of the sungrazing Kreutz comet C/2010 E6 (STEREO) in March 2010 at heliocentric distances $3-28~R_{\odot}$ were investigated using a newly-created set of analysis routines. The comet fully disintegrated during its perihelion passage. Prior to that, a broadening and an increase of the intensity peak with decreasing heliocentric distance was accompanied by a drop to zero polarisation at high phase angles (${\sim}105-135^{\circ}$, \textit{STEREO-B}) and the emergence of negative polarisation at low phase angles (${\sim}25-35^{\circ}$, \textit{STEREO-A}). Outside the near-comet region, the tail exhibited a steep slope of increasing polarisation with increasing cometocentric distance, with the slope becoming less prominent as the comet approached the Sun. The steep slope may be attributed to sublimation of refractory organic matrix and the processing of dust grains, or to presence of amorphous carbon. The change in slope with proximity to the Sun is likely caused by the gradual sublimation of all refractory material. The polarisation signatures observed at both sets of phase angles closer to the comet photocentre as the comet approached the Sun are best explained by fragmentation of the nucleus, exposing fresh Mg-rich silicate particles, followed by their gradual sublimation. The need for further studies of such comets, both observational and theoretical, is highlighted, as well as the benefit of the analysis routines created for this work.

\end{abstract}




\section{Introduction}

Prior to the turn of the century, most comet observations were limited to the observable night sky, i.e. at significant solar elongations. This prevented observations of most near-Sun comets -- which \citet{GeraintEtAl2018} define as having a perihelion closer than Mercury's perihelion distance; 0.307 AU or $66.1~R_{\odot}$ -- in their near-Sun regime. The advent of spaceborne solar observatories changed our picture completely: the \textit{SOHO} \citep[Solar and Heliospheric Observatory, ][]{SOHO1995} spacecraft -- and particularly its coronagraph instruments within the LASCO suite \citep[Large Angle Spectrometric Coronagraph,][]{LASCO1995} -- has at the time of this writing discovered over 4000 new comets\footnote{Featured article by S. Frazier for NASA Goddard Space Flight Center: \url{https://www.nasa.gov/feature/goddard/2020/4000th-comet-discovered-by-esa-nasa-solar-observatory}}, most of them near-Sun comets. The near-Sun environment imposes new extreme conditions on comets via insolation, solar winds, tidal forces, and sublimative torques, all of which can cause disruption to the nucleus directly, and will affect the dust properties directly or indirectly. The sublimative torques, specifically, can rapidly spin-up the nucleus and cause nucleus disintegration at those distances \citep{Jewitt1997}. It is therefore not surprising that most near-Sun comets break up near perihelion.

While the composition of comet volatiles has been a focus of extensive observations and study over the decades -- since spectroscopic observations of the coma and ion tail allow for the identification of the volatile species -- the dust composition and structure have remained more elusive. The two space missions which collected cometary dust in-situ and analysed it were \textit{Stardust}, which was a sample return mission \citep[e.g.][]{BrownleeEtAl2004}, and \textit{Rosetta}, where the samples were analysed locally, specifically with MIDAS and COSIMA instruments \citep[e.g.][]{RotundiEtAl2015,LangevinEtAl2016}. From them we have learned that the dust can be described by either compact or fluffy (highly porous) fractal aggregate particles, ranging in size from tens of micrometres to millimetres, all composed of smaller subunits \citep{MannelEtAl2016}. Starting with missions to comet 1P/Halley \cite{LawlerBrownlee1992}, we also learned that cometary dust is composed mostly of various silicate particles and an organic matrix component, though proportions vary between particles and, indeed, comets \cite{EngrandEtAl2016}. Notably, the returned samples from comet 81P/Wild 2 taken by the \textit{Stardust} spacecraft lacked a significant organic component \citep[e.g.][]{BrownleeEtAl2006,IshiiEtAl2008}, but remote observations of the coma showed a distinct presence of organics \cite{KisselEtAl2004}. The lack of organics in the samples may be attributed to the high-speed collisions with the aerogel during the collection rather than being an intrinsic property of the comet \cite{Brownlee2014}. Alternatively, the collection of material may have occurred in an unrepresentative area of the coma of comet, as investigated by \citet{ZubkoEtAl2012}. 

Since scattered light is polarised and the precise properties of polarisation depend on the properties of the scattering material -- in this case, cometary dust -- observation and analysis of polarised light can bring more constraints to our understanding of dust particle structure and composition \citep[e.g.][]{BagnuloEtAl2006, PolarComets, Levasseur-RegourdEtAl2018, HalderGanesh2021}. Most remote polarimetric observations of comets are concerned with measuring the variation of polarisation of the comet coma with phase angle or wavelength, from which the general polarimetric behaviour of comets has been determined \citep{PolarComets}, though occasionally change in polarisation along the tail is also considered \citep[e.g.][]{HadamcikEtAl2010,BorisovBagnuloEtAl2015,HadamcikEtAl2016,RosenbushEtAl2017,OleksandraEtAl2019}. Theoretical modelling of cometary dust has been able to successfully model the observed cometary behaviour so far, with a mix of silicates and refractory organics \citep[e.g.][]{KolokolovaJockers1997,KimuraEtAl2006,Kolokolova2016,KolokolovaEtAl2018,FrattinEtAl2019,ZubkoEtAl2020}. The effects of the near-Sun environment on the polarimetric properties of cometary dust, however, have rarely been explored before, although more general treatments of expected or observed physical and photometric behaviour have been made \citep[e.g.][]{Sekanina2000a,KimuraEtAl2002,SekaninaChodas2012}. In practice, most comet polarimetric analysis of comets to date has been determined for comets beyond the orbit of the Earth and observed from the Earth. Due to geometric considerations, this generally limits the maximum observable phase angle $\phi$ to $ < 90^{\circ}$. Near-Sun comets observed close to their perihelion do not suffer from this limitation, and can therefore provide us with glimpses into the high-$\phi$ region.

The results presented in this work show that polarimetry is a useful tool with which we can probe the effects of the near-Sun environment via the behaviour of refractory material in cometary dust. The results stem from \textit{STEREO} observations of sungrazing Kreutz comet C/2010 E6 (STEREO). Sungrazers are a subset of near-Sun comets, with perihelion distance $q$ between $1$ and $3.45~R_{\odot}$ ($0.0046-0.016$~AU) \citep{GeraintEtAl2018}. Comet C/2010 E6, as described in Section \ref{sec:obs}, is a typical member of the Kreutz family \citep{Kreutz1888,Kreutz1891,Kreutz1901}, the largest known family of comets, all members of which are sungrazers \citep{Marsden1967,Marsden1989,Marsden2005,BattamsKnight2016}. It was discovered in STEREO imagery a few days before its perihelion \citep{MPC_circular_2010IAUC}, from which it did not emerge, meaning it was likely (like most Kreutz comets) destroyed by the encounter. It was chosen for this study for its relative brightness, its typical orbital properties, and for its particularly clear photometric and polarimetric variability with changing heliocentric distance. Kreutz comets are observed primarily by the \textit{SOHO} and \textit{STEREO} spacecraft, all of which have some polarimetric capabilities. Their datasets are therefore a valuable reservoir of polarimetric observations of comets, and particularly for study of the effects of the near-Sun environment. 

A new method for data analysis is introduced, inspired by \cite{Thompson2015}'s analysis of pre-perihelion \textit{STEREO} and \textit{SOHO} spacecraft observations of comet C/2011 W3 (Lovejoy), and his later work on comet C/2012 S1 (ISON) \citep{Thompson2019paper}. The major difference between the approach to analysis in those works and here is that the spatial position of the comet is here determined from its orbital parameters, whereas \citet{Thompson2015} uses simultaneous observations by both \textit{STEREO-A} and \textit{B} spacecraft (described in Sec. \ref{sec:obs}) to determine the comet's position. The approach presented here allows for analysis of the comet even when only one spacecraft observes it, and it therefore has wider applicability. On one hand, this is very useful because it extends our window of observation: a longer time-series in the rapidly changing near-Sun environment helps draw more concrete conclusions, as exemplified in this work. On the other hand, since \textit{STEREO-B}'s loss of contact on 1st October 2014\footnote{More information available here: \url{https://stereo.nascom.nasa.gov/behind_status.shtml}}, only \textit{STEREO-A} has been sending data back to Earth. All near-Sun comet observations since that date present a perfect application of the new method, being independent of simultaneous observations.

The comet, observatories, and their imaging properties are outlined in Section \ref{sec:obs}. An overview of the image analysis procedure is presented in \ref{sec:method}, and the results in Section \ref{sec:res}. Their implications are discussed in Section \ref{sec:disc} and the findings summarised in Section \ref{sec:concl}.

\section{Observations}
\label{sec:obs} 

The comet discussed in this work was imaged by the twin \textit{STEREO} (Solar Terrestrial Relations Observatory) spacecraft; \textit{A} (Ahead) and \textit{B} (Behind). Launched in October 2006, the spacecraft sit in heliocentric orbits just inside (\textit{STEREO-A}) and outside (\textit{STEREO-B}) the orbit of the Earth \citep{STEREO2008}.

\begin{figure}
\centering
\includegraphics[width=1.0\columnwidth]{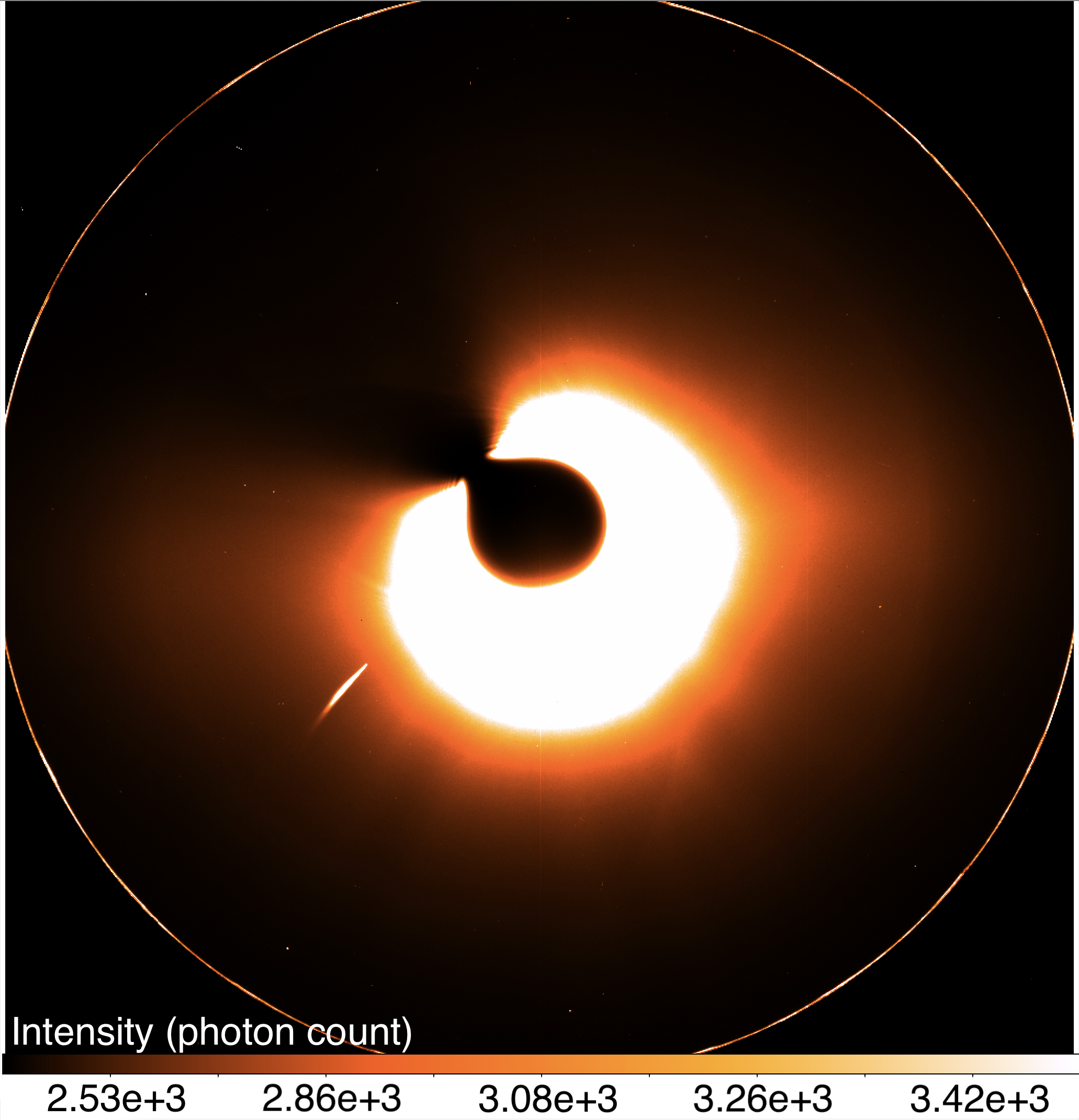}
\caption{\small Comet C/2010 E6 (bottom left) observed in the unprocessed field of \textit{STEREO-B}/SECCHI/COR2 camera on 12th March 2010 at 16:08:15 UT. Squared scaling is used to enhance the comet features. }
\label{fig:CK10E060_example}
\end{figure}

The equipment on the two spacecraft includes a set of coronagraphs in the SECCHI (Sun Earth Connection Coronal and Heliospheric Investigation) instrument suite \citep{SECCHI2008,BrewsherEtAl2010}. In this work only COR2 visible light (bandpass $650-750$ nm) coronagraph imagery is used. Its field of view, occulted in the centre, is between 2 and 15 solar radii at the Sun ($0.5 - 4.0^{\circ}$), and its CCD size is 2048x2048 pixels, with resolution of 15 arcsec/pixel; see Figure \ref{fig:CK10E060_example}. Like all coronagraphs, it suffers from some vignetting near the occulted region. Lack of different bandpass filters prevents a multi-wavelength analysis of the data.

Permanently in the light path of COR2 is a linear polariser which can be set at three rotation angles: $0^{\circ}$, $120^{\circ}$, and $240^{\circ}$ relative to a reference position. A sequence (triplet) of images using the three different polariser angles in quick succession is taken either twice (2006-2009) or once (2009-) per hour, with other observational procedures (generally using rotation angle of $0^{\circ}$ only) taking place the rest of the time. For observations discussed here, the polarimetric triplet imaging sequence was conducted hourly with images of the triplet taken in quick succession at $8'15''$, $8'45''$, and $9'15''$ UT past the hour, with an average exposure time of 6 seconds. This procedure was identical for both spacecraft.

The comet observed and presented here is C/2010 E6 (STEREO). It belongs to the Kreutz family (or group) of sungrazing comets, meaning comets with perihelion distance below $3.45~R_{\odot}$ \citep{GeraintEtAl2018}. Kreutz comets are the largest known family of comets, representing $86~\%$ of SOHO comet observations  \citep{BattamsKnight2016}. They share extreme orbital characteristics, namely high eccentricity, small perihelion distance, and high orbital inclination, though some variations within the group exists, leading to a common differentiation into two groups \cite{Marsden1967,Marsden2005,SekaninaChodas2004}. Due to highly elliptical trajectories and short observation windows, their eccentricity is often assumed to be $e =1$. 

The average size of a Kreutz comet is estimated at $<100$~m, with the smallest observed members at $5-10$~m \citep{Sekanina2003}. They are thought to have originated from a single parent body which fragmented over the course of a few centuries or millenia, with the orbit moving closer to the Sun \citep{SekaninaChodas2004,SekaninaChodas2007}. This may be a typical dynamical end-state of comets \citep{BaileyEtAl1992}. The orbital parameters of Kreutz comets overall and of comet C/2010 E6 -- which had orbital parameters typical of its family -- are presented in Table \ref{tab:orb_para}. 

\begin{figure}
\centering
\includegraphics[width=1.0\columnwidth]{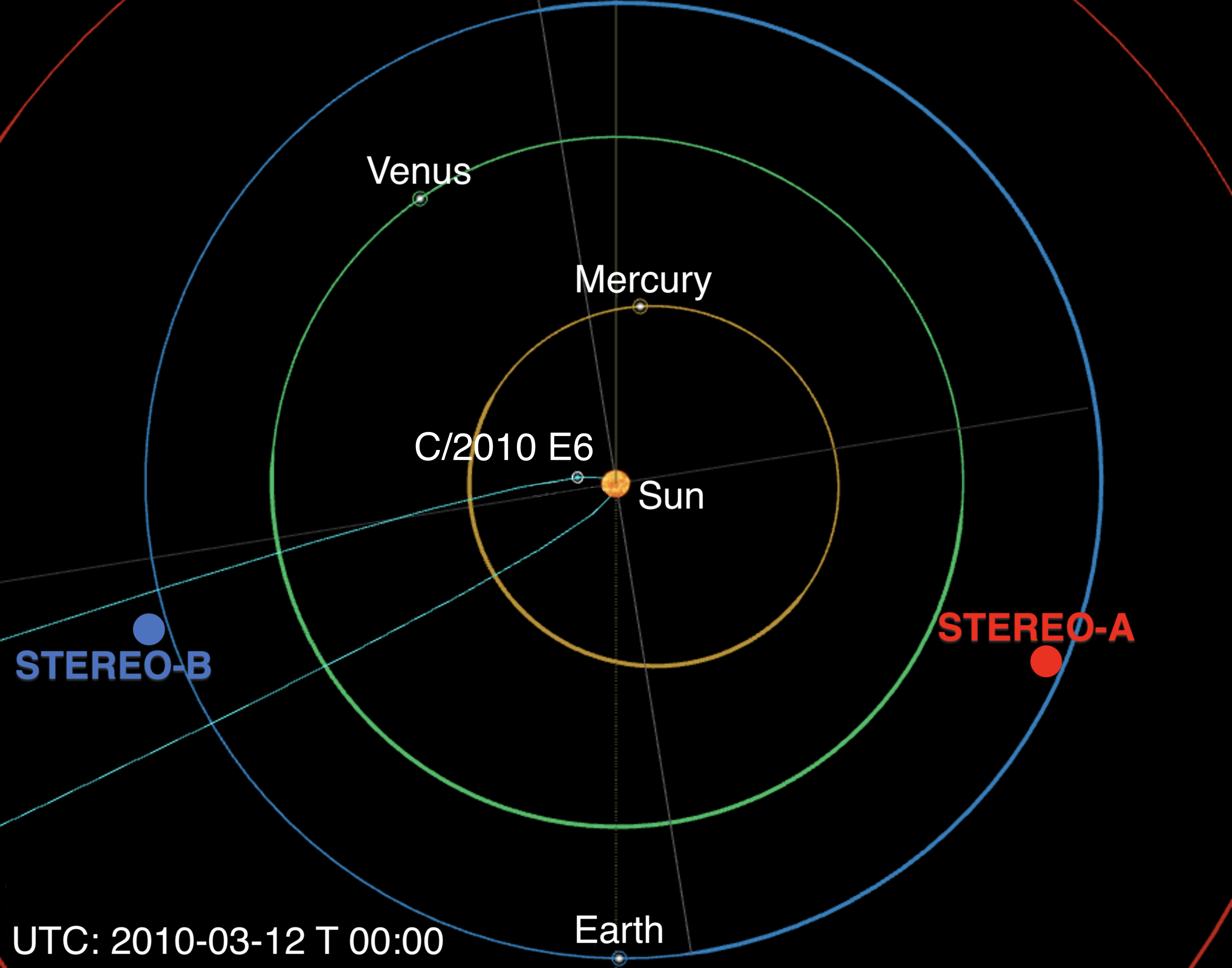}
\includegraphics[width=1.0\columnwidth]{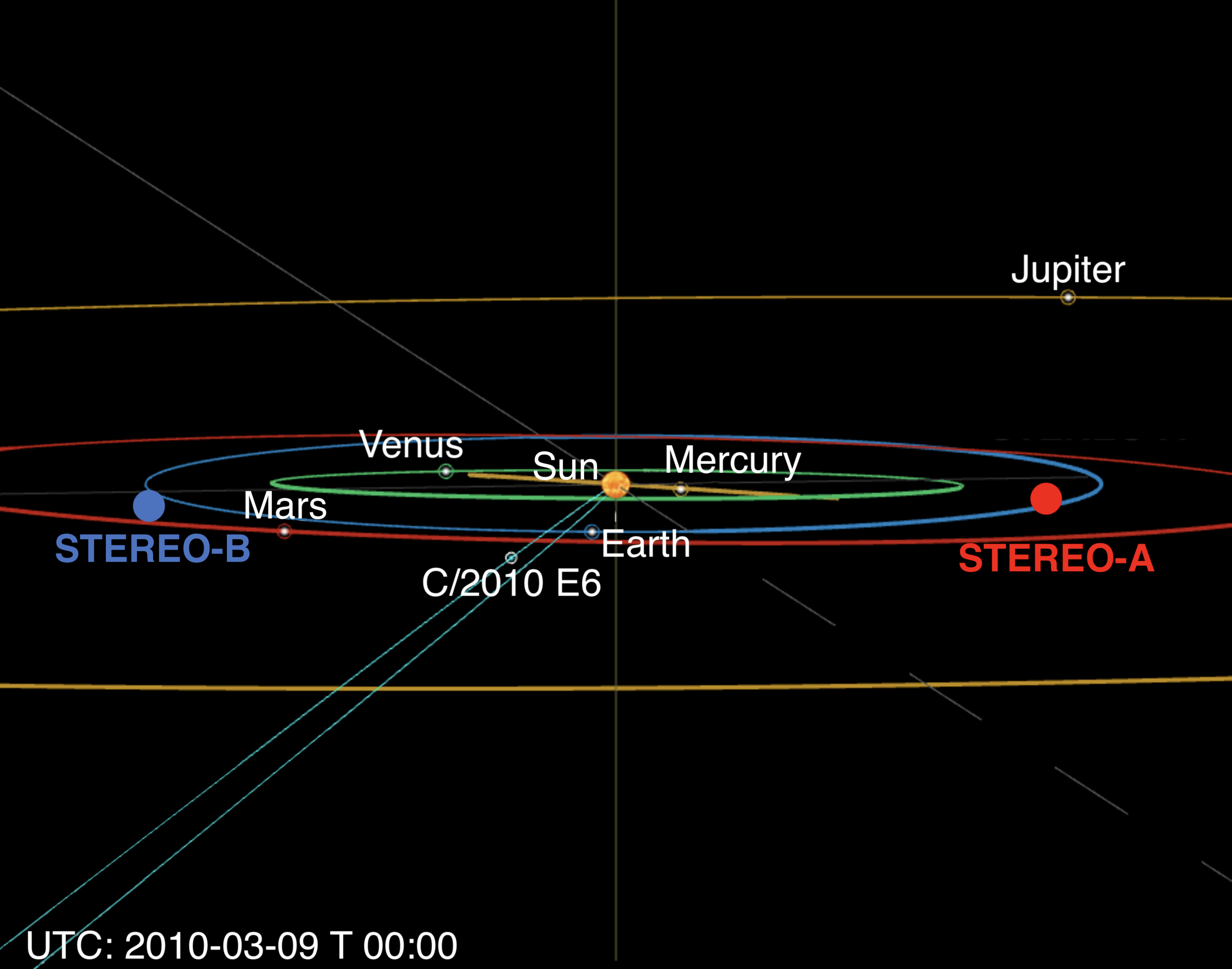}
\caption{\small Orbit of comet C/2010 E6 (STEREO) -- cyan line with comet location labelled -- and the positions of the \textit{STEREO} spacecraft in the context of the inner Solar System\protect\footnotemark. The top image is the view on 12th March 2010 at midnight UT from above the ecliptic plane. The bottom image is a `side-on' view from close to the ecliptic plane, with the Earth-Sun line from 12th March aligned with the Sun's $z$ axis (vertical line in the centre of the image), but with imagery shifted to 9th March 2010 to render the comet label readable. \textit{STEREO-A} (red filled circle) and \textit{B} (blue filled circle) positions were superimposed onto the orbital image and are approximate\protect\footnotemark. }
\label{fig:CK10E060_AB_orbits}
\end{figure}

Virtually all Kreutz comets are observed only prior to perihelion and presumably fail to survive it due to their small size and small perihelion distance. The same is true for comet C/2010 E6. Extrapolating from the theory of Kreutz group formation, it is presumed that the comet fragmented from the parent near its previous perihelion, making this its only perihelion passage as a distinct object. Additionally, the comet has both a faint precursor, C/2010 E10 (SOHO) and a pair of faint successors -- C/2010 E11 and E12 (SOHO) \cite{2010MPECRuanEtAl} -- the three companion fragments are separated from the bright comet E6 by about a day on either side. Such observations have precedents: several near-Sun comets have been observed arriving in pairs or larger clusters within a few days of one another \cite{KnightEtAl2010}. The main theory describing them posits that they are a result of non-tidal, secondary fragmentations occurring at large heliocentric distances \citep{Sekanina2000a,Sekanina2002}. It may be of interest to compare their photometric and polarimetric properties to that of comet E6 in the future, though it is possible they are not closely related, beyond simply belonging to Kreutz family. The orbit of the comet in the context of the inner Solar System from two different vantage points as well as the locations of the twin \textit{STEREO} spacecraft at the time of these observations are presented in Figure \ref{fig:CK10E060_AB_orbits}.

\footnotetext[7]{Using `Catalina Sky Survey Orbit View' by D. Rankin: \url{https://catalina.lpl.arizona.edu/css-orbit-view} and the comet orbital information from the Minor Planet Center.}
\footnotetext[8]{Using `Where is STEREO' tool: \url{https://stereo-ssc.nascom.nasa.gov/where.shtml}}

Comet C/2010 E6 (STEREO) -- discovered by \textit{STEREO} -- was seen in both \textit{STEREO-A} and \textit{B} SECCHI/COR2 fields of view, in the former from $2010/03/11$ at 15:08 UT (heliocentric distance of nucleus $r =26.8~R_{\odot}$) to $2010/03/12$ at 19:08 UT ($r=6.3~R_{\odot}$), in the latter from $2010/03/11$ at 23:08 UT ($r=21.7~R_{\odot}$) to $2010/03/12$ at 21:08 UT ($r=3.5~R_{\odot}$), which includes 18 hours of simultaneous observations with both spacecraft. As discussed in Section \ref{sec:res}, the comet coma underwent significant brightening as it approached the Sun; from $2.3\cdot 10^{-10}~L_{\odot}$ to $4.5\cdot 10^{-8}~L_{\odot}$ at its peak, a near fifty-fold increase in brightness in 23 hours (from $2010/03/11$ at 15:08 UT to $2010/03/12$ at 14:08 UT). Due to the comet's proximity to the Sun, the latter cannot be trivially assumed to act as a point source; the Sun subtends an angle of $4.3^{\circ}$ at the largest heliocentric distance of the nucleus observed in this work ($26.8~R_{\odot}$, as above), and an angle of $31.9^{\circ}$ at the smallest observed heliocentric distance ($3.5~R_{\odot}$). This effect is explored further in Sections \ref{sec:res} (Results) and \ref{sec:disc} (Discussion).

The comet was also observed by the \textit{SOHO}/LASCO C2 and C3 coronagraphs. \textit{SOHO}/LASCO coronagraphs are arranged differently from \textit{STEREO}/SECCHI ones. Most notably, unlike the latter, the former includes three distinct polarisers at three different rotation angles ($0^{\circ}$, $120^{\circ}$, and $240^{\circ}$) mounted on a filter wheel alongside a clear glass position (no polariser) and another filter \citep{LASCO1995}. This means, firstly, that a polariser is not permanently in the light path, and a full set of polarimetric images is generally only taken once per day (C3) or $3-4$ times per day (C2). Secondly, the $0^{\circ}$ polariser of the C3 coronagraph has been out of commission for most of the \textit{SOHO} mission lifetime, though workarounds using the clear glass image may be used to compensate for that \citep[e.g.][]{GrynkoEtAl2004,Thompson2015}. Thirdly, \textit{SOHO} mission is in part a more prolific comet discoverer than \textit{STEREO} due to the wider bandpasses on the coronagraphs which include the NaI doublet \citep{BieseckerEtAl2002}. That is strong in many near-Sun comets, making them brighter than in \textit{STEREO} imagery. The inclusion of strong gas signatures in SOHO filters, however, may contaminate polarimetric signal from the cometary dust. \cite{Thompson2015} has shown that these drawbacks make \textit{SOHO} data much less useful for discussion of polarimetric properties. Comet C/2010 E6 passed very near the support for the occulter disk for the C3 coronagraph, making photometric data less reliable as well. \textit{SOHO} data analysis of the comet is therefore not included in this work.

\begin{table}
\centering
\caption{Table of average orbital parameters of Kreutz family \citep{BattamsKnight2016} and comet C/2010 E6 (STEREO) \citep{MPC_circular_2010IAUC}.}
\label{tab:orb_para}
\begin{tabular}{l|r|r} 
	Orbital parameter & Kreutz & C/2010 E6 \\
    \hline
	Orbital inclination $i$ [$^{\circ}$]  & $143.2$ & $144.60$ \\
	Longitude of the ascending node $\Omega$ [$^{\circ}$] & $0.4$ & $4.381$ \\
	Argument of perihelion $\omega$ [$^{\circ}$] & $80.0$ & $83.206$ \\
	Eccentricity $e$ & $>0.9999$ & $1.0$ \\
    Distance of Perihelion $q$ [AU] & $0.0056$ & $0.00480$ \\
    $~~~~~~~~~~~~~~~~~~~~~~~~~~~~~~~~~~~~~~~~[R_{\odot}]$ & $1.2$ & $1.0317$ \\
    Date of Perihelion Passage & $n/a$ & $2010/03/12$ \\
    Time of Perihelion Passage (UT) & $n/a$ & 21:26 \\
	\hline
\end{tabular}
\end{table}

\section{Data Reduction}
\label{sec:method} 

The image analysis and data plotting were both conducted using a bespoke IDL (Interactive Data Language\footnote{IDL version 8.7, Harris Geospatial Solutions, Boulder, Colorado}) package named \texttt{composite} (\textbf{com}et \textbf{po}larisation \textbf{s}ystematic \textbf{i}ntegration-based analysis \textbf{te}chnique). It is a semi-automated series of routines fine-tuned for use with \textit{STEREO}/SECCHI/COR2 polarised image triplet data. It was inspired by similar work on \textit{STEREO} observations of comet C/2011 W3 (Lovejoy) \citep{Thompson2015} and, more recently, comet C/2012 S1 (ISON) \citep{Thompson2019paper}.

\begin{figure}
\centering
\includegraphics[width=1.0\columnwidth]{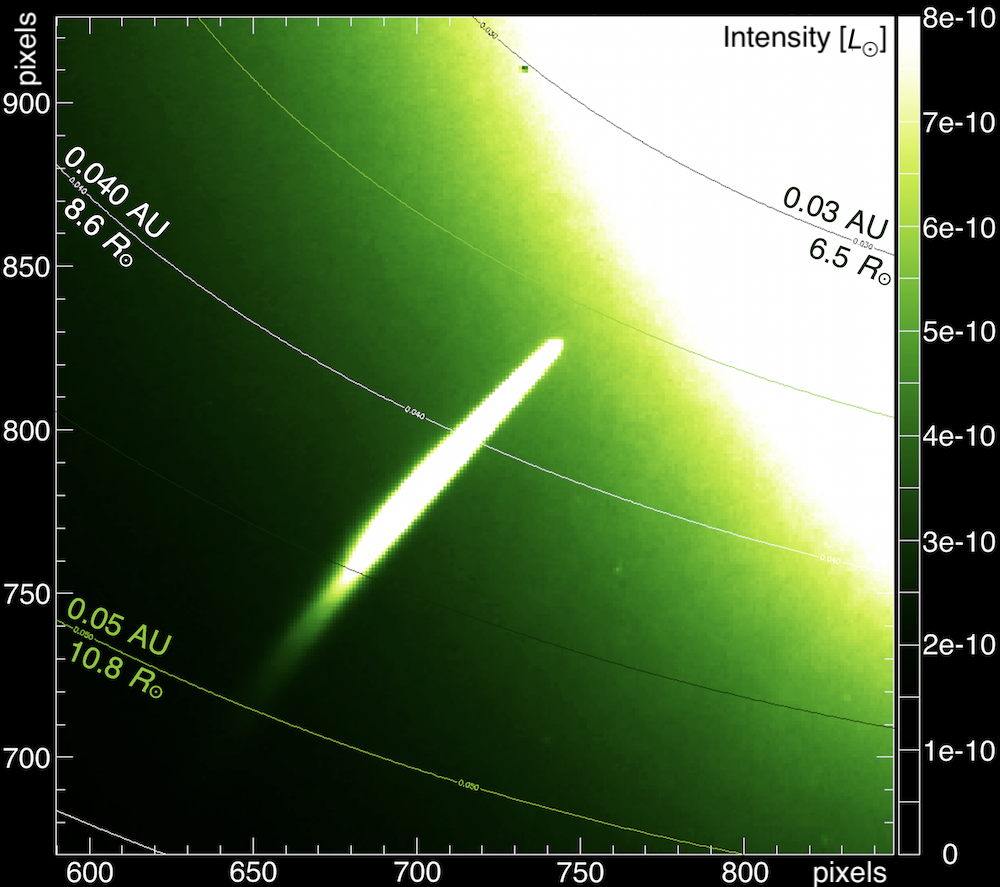}
\caption{\small Close-up view of comet C/2010 E6 observed in the field of \textit{STEREO-B}/SECCHI/COR2 camera on 12th March 2010 at 16:08:15 UT with colour gradient in units of solar luminosity, and contour lines showing the orbital plane distance to the Sun in AU. The $x$- and $y$-axes are pixel counts for the full image, starting in the bottom left corner.}
\label{fig:CK10E060_sundist}
\end{figure}

Initially the image triplets undergo standard pre-processing using the \texttt{SECCHI\_PREP} routine from the SolarSoft library for IDL -- a handful of other useful SolarSoft routines is utilised within \texttt{composite} as well \citep{EichstedtThompsonEtAl2008,SolarSoft1998}. Orbital parameters for the comet from the relevant Minor Planet Center circular are utilised \citep{MPC_circular_2010IAUC}. Along with the positional and pointing information of the spacecraft, they are used to find the plane of the comet's orbit, in which the dust tail is assumed to lie; all data points in the images are then mapped onto this plane. From this the distances between the comet (and points along the tail) and the Sun, the spacecraft, etc. can be determined, as well as all the relevant angles, including the phase angle $\phi$ between the Sun, the comet and its tail, and the spacecraft. Some of the extracted data is overplotted on the coronagraph imagery in Figure \ref{fig:CK10E060_sundist}.

\begin{figure}
\centering
\includegraphics[width=1.0\columnwidth,trim={20.2cm 1.0cm 6.2cm 4.6cm},clip]{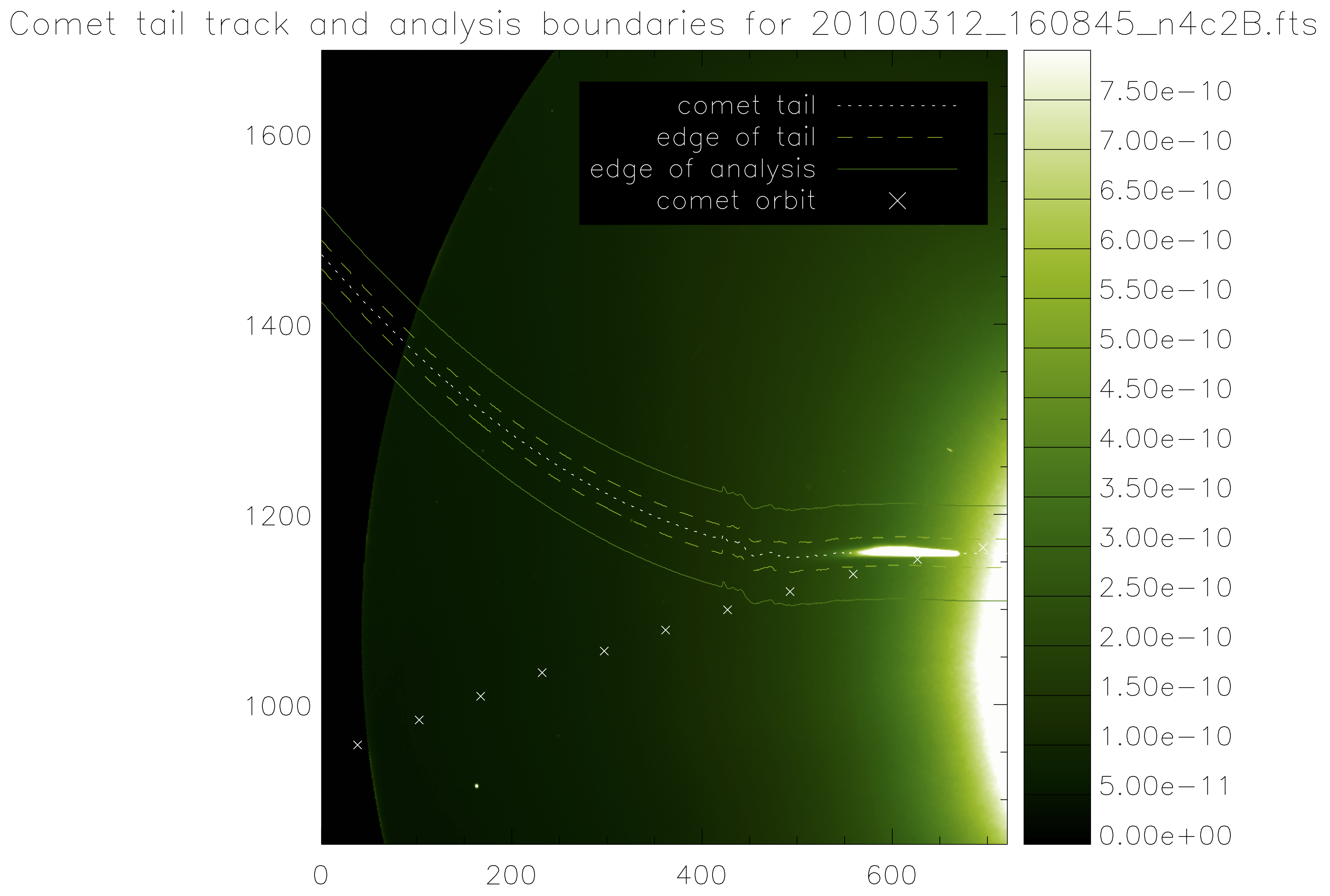}
\caption{\small Close-up view of comet C/2010 E6 observed in the field of \textit{STEREO-B}/SECCHI/COR2 camera on 12th March 2010 at 16:08:15 UT with colour gradient in units of solar luminosity, showing the tail tracing procedure. `Edge of analysis' refers to the edge of the cross-section region for which the cubic fit is determined; outer limits of Fig. \ref{fig:CK10E060_crosssec}. The $x$- and $y$-axes are pixel counts for the full image, starting in the bottom left corner. Comet photocentre is positioned at $x$-axis pixel value of 675.}
\label{fig:CK10E060_tailtrace}
\end{figure}

Then, for the full time interval in which the comet is seen in the coronagraph , the comet photocentre (the intensity peak in the coma region) is found interactively in the image triplets. The comet tail is then traced in each separate image using two different methods -- the first using detailed tracing the local brightness peak, the second interpolating a smoother trace from the first -- and a visual inspection of the tail tracing performed as a quality check. Without that quality check, background objects and image artefacts may dominate the tail tracing direction. Points beyond the distinguishable tail region will be dominated by them, and are discarded in the final step of data processing; e.g. it is clear in Figure \ref{fig:CK10E060_tailtrace} that traces more than ${\sim}250$px to the left of comet photocentre trace the background image rather than the -- by then nonexistent -- tail. The image is rotated so that the pixels along the tail form a near-horizontal line for easier creation of cross-sections. The successive transverse (vertical) cross-sections are found, starting ahead of the comet photocentre and progressing along the tail (or, if the tail is short, tracing the background signal) until the edge of the image. 

\begin{figure}
\centering
\includegraphics[width=1.0\columnwidth,trim={0.7cm 0.9cm 1.3cm 2.1cm},clip]{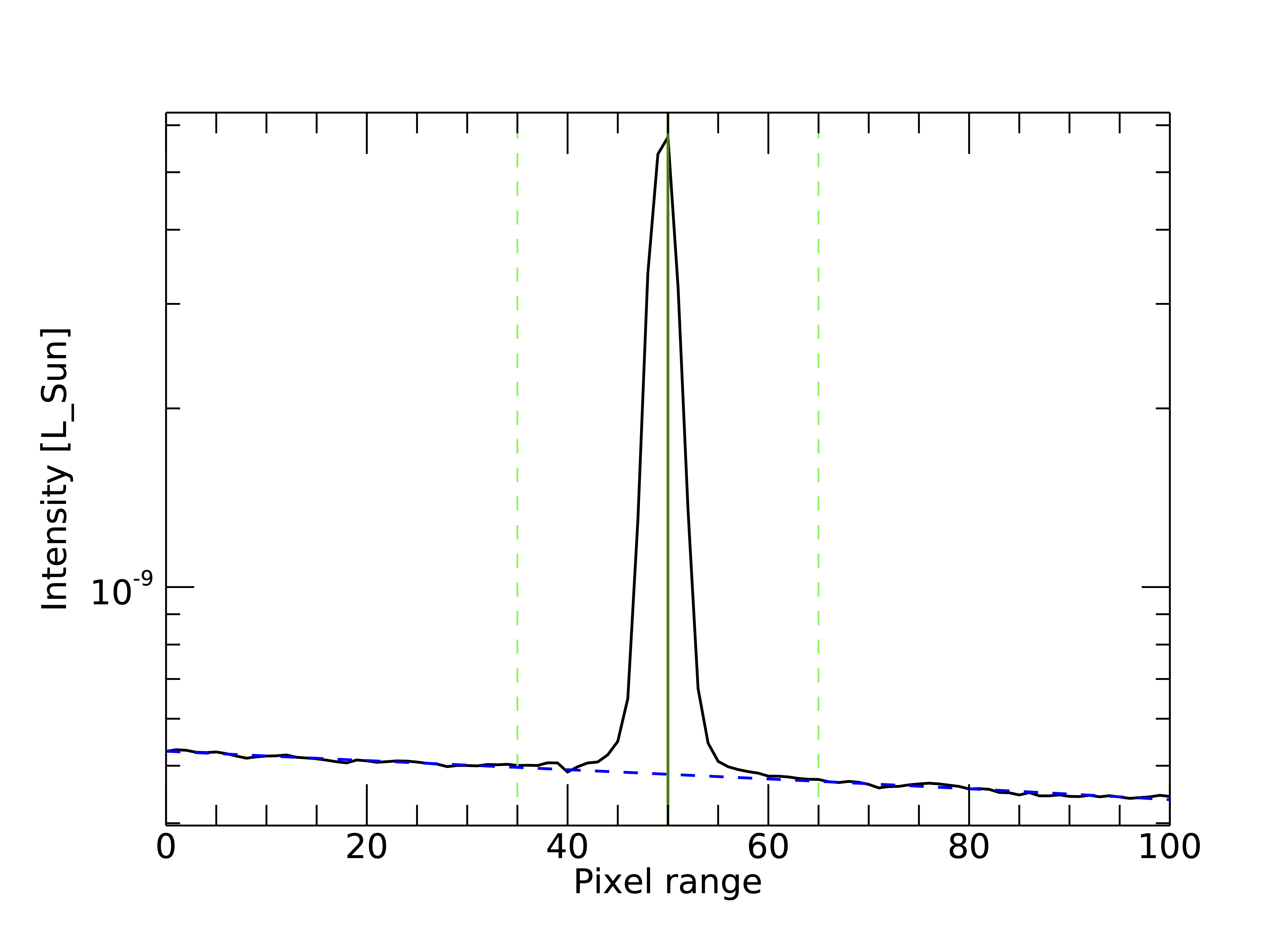}
\caption{\small An example cross-section (10th consecutive one away from comet photocentre) of comet C/2010 E6 observed in the field of \textit{STEREO-B}/SECCHI/COR2 camera on 12th March 2010 at 16:08:15 UT. The vertical  dark green solid line in the centre denotes the peak brightness of the comet tail cross-section, while the vertical light green dashed lines on either side of it show the lateral extent of the `tail region'. The near-horizontal dark blue dashed line is tracing the cubic fit for background intensity.}
\label{fig:CK10E060_crosssec}
\end{figure}

Being traced along the comet tail, the cross-sections are centred on the intensity peak caused by the comet; see Figure \ref{fig:CK10E060_crosssec} for an example. A central `tail region' is reserved and a local cubic background fit determined for each cross-section, excluding the `tail region'. The limits of the `tail region' and the cubic fit region depend on the brightness profile of the comet; the former may generally be 20-30 pixels (10-15px on each side of the peak) and the latter 60-100 pixels (30-50px on each side of the peak). They are determined based on the maximum width of the comet signal in the full imagery dataset: for comet C/2010 E6 they have been chosen as tail region of  20 pixels and cubic fit region of 60 pixels in \textit{STEREO-A} data, and 30 pixels and 100 pixels, respectively, in \textit{STEREO-B} data due to the differences in tail brightness stemming from different geometry of observations (see Fig. \ref{fig:CK10E060_AB_orbits}).

High quality background subtraction is challenging in the coronagraph imagery of the highly spatially and temporally variable near-Sun environment, and multiple background fitting options were considered. Using images taken within a few hours of the target image as reference is not suitable for this application, as the comet takes several hours to pass across the field of view. Due to the high orbital inclination of Kreutz group comets the curve of their dust tails -- observed from the ecliptic plane -- usually closely follows their orbit. Thus a `background' image will itself likely include the comet overlapping the `foreground' comet imagery, corrupting the results. Alternatively, an image prior to the comet's appearance in the field of view may be used, but the variability of the near-Sun environment on the scale of hours is significant enough to make that option unreliable. A cubic fit is instead applied to each cross-section to remove the background (see Fig.~\ref{fig:CK10E060_crosssec}), and the `tail region' truncated (integrated) to a single step for each cross-section. This is done in order to boost the signal, since the image resolution is invariably too poor for tracing two-dimensional variability. The process is repeated for each longitudinal step of 1 pixel, and for each image in the triplet.

The three resulting polarised intensity vectors from the three orientations of the polariser in the image triplet -- $I_{0}$, $I_{120}$, and $I_{240}$ (for the orientations of $0^{\circ}$, $120^{\circ}$, and $240^{\circ}$, respectively) -- are then coarsely aligned with each other based on the orbital information, with additional fine-tuning minimising the differences of the three curves. The movement of the comet in space in the 30 seconds between each image in the triplet has a negligible effect, however the comet can move by up to 2 pixels as the filter is rotated, likely due to optical effects. This prevents a simpler analysis method where the full coronagraph images could be analysed at once by simple stacking, without explicitly identifying and tracing the comet in the field of view. Due to the narrowness of the comet tail, the sharp transitions between the tail and the background, and relatively low spatial resolution of the imagery, such simplified stacking produces highly distorted results dominated by artefacts and is not utilised here. Furthermore, a correction value $A$ must be included to factor in geometric offset effects. It takes into account the position angle of the scattering plane with respect to a chosen reference point, as well as instrumental effects, in particular potential offsets of the polarisers from that reference point. 

From here the Stokes parameters defining the intensity (Stokes $I$) and the degree of linear polarisation -- reduced Stokes $Q$ and $U$ ($Q/I$ and $U/I$) -- can be calculated along the tail as follows, with $\alpha \in \{0, 120, 240\}$:
\begin{equation}
\label{eq:I}
I = \frac{2}{3}\left(\sum\limits_{\alpha}I_\alpha\right)
\end{equation}

\begin{equation}
\label{eq:QIalter2}
\frac{Q}{I} = \frac{2\sum\limits_{\alpha}I_\alpha \cos(2(A - \alpha))} {\sum\limits_{\alpha}I_\alpha}
\end{equation}
\begin{equation}
\label{eq:UI}
\frac{U}{I} = \frac{-2\sum\limits_{\alpha}I_\alpha \sin(2(A - \alpha))} {\sum\limits_{\alpha}I_\alpha}
\end{equation}

The standard calculation of uncertainties assume that shot noise scales with the inverse of the signal-to-noise ratio, but the more pertinent measure of uncertainty is the degree of scatter in polarimetric plots. The uncertainties derived from the comet signal analysis are negligible for high signal-to-noise ratio, but slowly increase moving along the dimming comet tail, up to a few percent in polarisation at largest cometocentric distances. The uncertainties from aligning the images from the triplet are more difficult to quantify. They have been minimised by truncation (integration) of the tail cross-sections, increasing the signal-to-noise ratio, and by visual inspection of the alignments. The knock-on effects of potential misalignments will be the greatest in two regions: first at the photocentre of the comet, where small variations in the relative slopes of the polarised intensity curves can significantly affect the calculated polarisation, and second at large cometocentric distances, where low signal can cause large variations in absolute values of polarised intensity. Due to the relatively low resolution of observations, this effect cannot be fully avoided. The potential for errors inherent in the first section affecting the overall conclusions is discussed in Section \ref{sec:disc}. Since the calculated uncertainties are lower than the scatter of the data points, error bars are not included in the plots, with the degree of scattering better indicating the quality of the data.

\begin{figure}
\centering
\includegraphics[width=1.0\columnwidth,trim={1.2cm 0.7cm 1.5cm 1.9cm},clip]{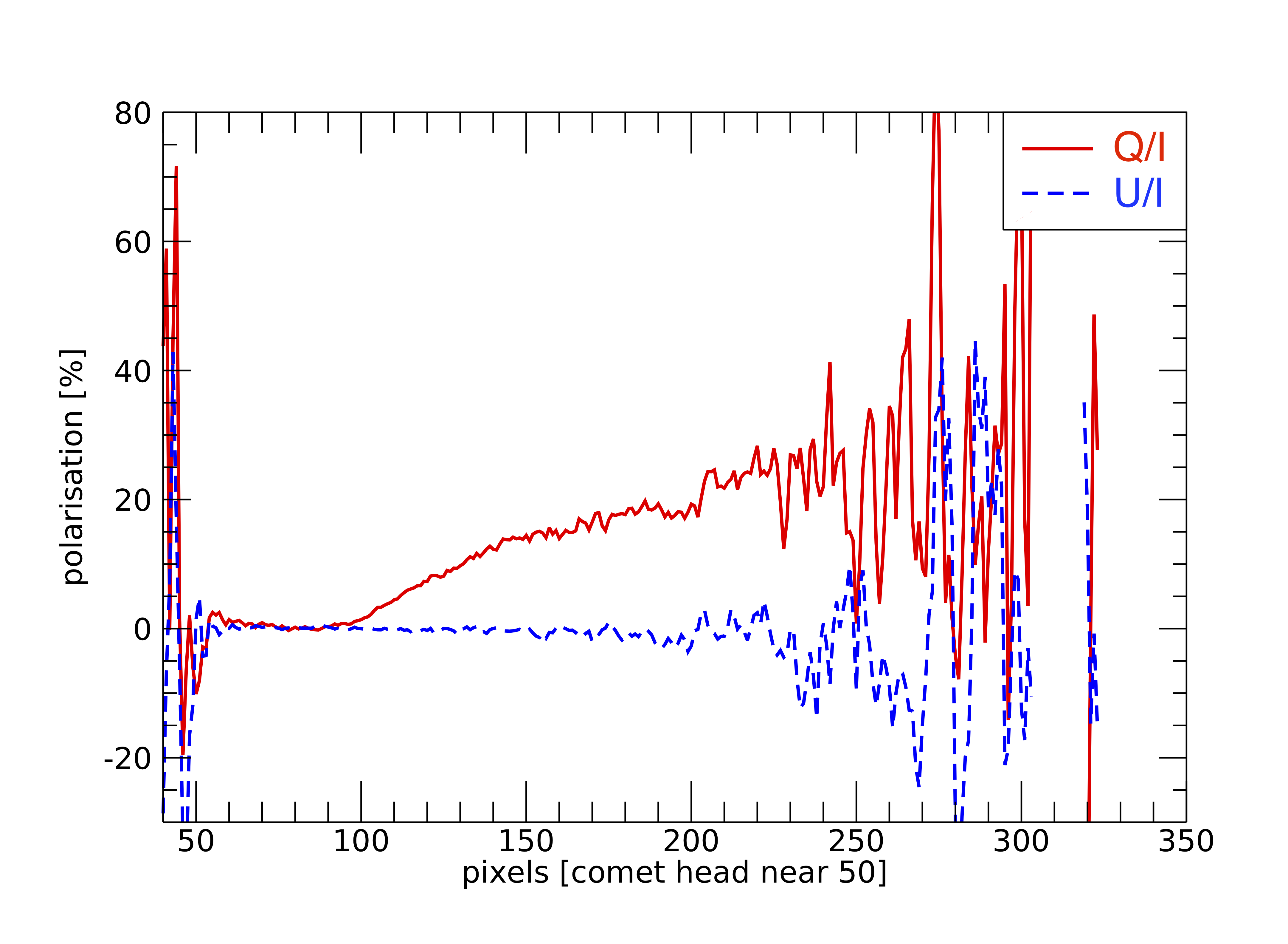}
\caption{\small A plot of Stokes parameters $Q/I$ and $U/I$ for comet C/2010 E6 observed in the field of \textit{STEREO-B}/SECCHI/COR2 camera on 12th March 2010 at 16:08:15 UT against the tail-tracing pixel values. The red full line denotes $Q/I$, the blue dashed line $U/I$. Comet photocentre sits near pixel 50 and pixel values increase along the tail with increasing heliocentric and cometocentric distance. Here ten pixels span a distance of approximately $0.19~R_{\odot}$ (${\sim}135,000$~km).}
\label{fig:CK10E060_QUP}
\end{figure}

The upper cut-off point for the plots is the edge of the tail as determined by a combination of visual inspection of tail tracing and the degree of scatter of Stokes parameters, as well as scatter of the total degree of linear polarisation $p$ (where $p^2 = (Q^2+U^2)/I^2$). Comparing the scatter in Figure \ref{fig:CK10E060_QUP} to the tail tracing in Figure \ref{fig:CK10E060_tailtrace}, we see the data beyond ${\sim}200$th pixel along the tail away from the comet photocentre (i.e. pixel values $>250$ there and the $x$-axis pixel values $<475$ in Fig. \ref{fig:CK10E060_tailtrace}) can be ignored, since the tail tracing method was derailed by either background signal or various artefacts as the signal-to-noise ratio along the comet tail deteriorated. Such data beyond the realistic observational limit of the comet tail is not used in any of the remaining analysis. 

The large calculated range of polarisation at the comet photocentre, seen in the following section, should be highlighted here. Some of this apparent large range (particularly in \textit{STEREO-A} data) is due to real variability along the near-nucleus tail, better seen in Figure \ref{fig:CK10E060plotsIQI}, whereas the sparse data points with highly negative polarisation seen especially in \textit{STEREO-B} data are artefacts of the polarisation calculations derived from low-resolution imagery, indicated in the previous section. The analysis procedure generally produces tail traces as smooth curves of related data points, which agrees with the theories of light scattering from evolving dust particles, and thus major deviations from those curves are likely to be artefacts. Figure \ref{fig:CK10E060_ABQ}.ii is included as a check for reliability of Fig. \ref{fig:CK10E060_ABQ}.i data: since $U/I$ should be zero: this is expected when the dust particles don't have a preferred orientation in space, and has been consistently born out by polarimetric observations of cometary dust thus far. Significant deviations from that will show a reduction in quality of the data rather than a change in physical characteristics.

\section{Results}
\label{sec:res} 

\subsection{Sun as an extended source }
\label{sec:extendedSun} 

Before we discuss the observational results in full, the effect of the Sun as an extended source must be investigated. For the data closest to the Sun (at heliocentric distance of $3.5~R_{\odot}$), the Sun subtends an angle of $31.9^{\circ}$ in the sky. This has an effect not only on the processes affecting the comet -- such as dissociation, ionisation, and sublimation, affected by the photon fluxes as well as the solar wind and local temperature \cite[e.g.][]{GeraintEtAl2018} -- but on geometric considerations like the phase angle and scattering plane of the incoming radiation.

The maximum variations in the phase angle and in the angle of the scattering plane from the expected (mean) value both equal half the angular size of the Sun in the sky and are perpendicular to one another. This varies from $2.1^{\circ}$ to $15.9^{\circ}$ (see Sec. \ref{sec:obs}). To consider the effects such a large spread of angles may have on the results of this investigation, a number of factors must be taken into account. Assuming the Sun is a perfect sphere -- effects of oblateness are minimal when compared to the effect of non-negligible angular size -- the sum of all light rays will average to the expected (mean) value of the phase angle, and to the scattering plane expected from a central point source. Thus the reported values of phase angle and polarisation remain valid as the mean value  characterising the observing geometry. Additionally, the fact that light from near-mean angles comes from a larger area than light from near the edges (due to the nature of a sphere), in combination with limb darkening, means a greater proportion of incoming light will have mean or near-mean values rather than values from near the edges. These factors will likely diminish the effect of the non-negligible size of the Sun on the results.

We have analysed the potential effect of the spread of scattering planes of incoming light on the observed Stokes parameters. Light from a scattering plane e.g. $15.9^{\circ}$ from the mean one -- used for calculations -- will act as light polarised at a $15.9^{\circ}$ angle to the mean orientation. Due to symmetry, this angled light will, when arriving from both sides of the Sun, reinforce in Stokes $Q$ calculations, but cancel out in Stokes $U$ calculations. This means the deviation cannot be traced using Stokes $U$ alone. Indeed, an investigation of Fig. \ref{fig:CK10E060_ABQ}.ii reveals no evidence of deviation: \textit{STEREO-B} data (higher phase angles) shows an increase in spread of $U/I$ with increasing phase angle, which correlates with increasing heliocentric distance and therefore fainter signal and is actually anti-correlated with the angular size of the Sun. 

Investigating the effect of Sun as an extended object on the observed Stokes $Q$, and of limb darkening in particular, we found that even at the closest heliocentric distances discussed in this work, the signal from the central regions will greatly dominate that from the limbs. For fully linearly polarised light, $Q/I$ from the limbs would register as ${\sim}0.848$ instead of $1$, and the effect scales with polarisation (i.e. the signal from that angle will always be diminished by that factor), but when the effect is averaged over the limb-darkened surface of the Sun, the average signal is scaled only by a factor of ${\sim}0.95$ compared to the true signal. The effect also rapidly diminishes with heliocentric distance: at $5~R_{\odot}$, the factor is $>0.97$, and at $10~R_{\odot}$ it is $>0.993$. It should be noted that limb-darkening calculations were done using results at $1$~AU; closer to the Sun the limb regions will be even darker, so the factors calculated above are lower estimates, and the real effects are likely even less prominent. This source of error, then, is generally smaller than other sources of uncertainty in this analysis, and can be safely ignored. It is, however, systematic in nature, so it could be accounted for in the results if the need arose. 

The effect of the Sun as an extended object on the phase angles of the light is more straightforward: if \textit{STEREO-A} observations at the smallest heliocentric distance ($\sim7~R_{\odot}$) were taken at the phase angle of $\phi \sim 25^{\circ}$, but the Sun subtends $16.3^{\circ}$ in the sky at that point, so the full range of phase angles of light reaching the coronagraph camera will be $\phi\sim 16.8^{\circ}$ to $\phi\sim 33.2^{\circ}$. The lower end of that range sits firmly within the standard negative polarisation branch observed for comets, which might go some way towards explaining the negative polarisation seen in the data (e.g. Fig. \ref{fig:CK10E060plotsIQI}). The higher end, however, sits firmly in the positive branch, so the two effects may cancel out, depending on the slope of the phase angle curve.

Furthermore, the same consideration must be applied here as for the scattering angle: off-centre contributions will have, proportionately, less of an effect than central ones. The inner tenth of the angle range contributes $\sim 13.6\%$ to the total intensity, while the outer tenth (both limbs combined) contributes only $\sim 2.9\%$ (using limb-darkening estimates at $1$~AU again, making the latter an upper limit). Scaling the deviation from the central phase angle with the relative amount of signal produced at those deviations, we find that the error for the phase angle may be quantitatively expressed as $\sim41\%$ of the angular radius of the Sun. For the example of \textit{STEREO-A} data from the previous paragraph, we thus have $\phi = 25^{\circ} \pm 6.7^{\circ}$. This error, like the angular size of the Sun, rapidly diminishes with distance. Due to peculiar behaviour of comet C/2010 E6 (STEREO), it is unclear if this spread of phase angles has any significant effects on the observations.

Thus, the effect of the Sun as an extended light source on phase angle and polarisation cannot be easily ascertained from the data presented here. No significant effect is produced from the scattering angle variation, and the phase angle variation, while notable on its own, might not have a significant effect on the results unless the phase angle curve is not smoothly increasing or decreasing. This may be fruitful ground for future investigations, however. Ideally the same comet would be observed at the same phase angle twice or more: once when the comet is at some distance from the Sun, where it may be treated as a point source (e.g. at $1$~AU or above, where it subtends $\leq 0.5^{\circ}$), and once much closer to the Sun, with its angular size at several degrees. Then the difference in the polarisation signatures could begin to be attributed to the broadening of the phase angle signals in particular. Even then the effects of the near-Sun environment during the second observation would need to be taken into account first, so a near-Sun comet with known properties like 96P/Machholz might be a good candidate instead of a Kreutz comet. It is possible, though unlikely, that a comet has already been fortuitously observed in this manner: the STEREO and SOHO spacecraft tend to only observe comets when they are very close to the Sun, while most near-Sun comets have not been observed beyond $1$~AU, especially not with polarimetric equipment. Comet C/2012 S1 (ISON) might have been a good candidate, as the Hubble Space Telescope made polarimetric observations of it at $3.81$~AU, but the phase angle was $12.16^{\circ}$ \citep{HinesEtAl2014} while the Sun-observing spacecraft only saw the comet at higher phase angles within their coronagraphs \cite{Thompson2019paper}. The Vera C. Rubin Observatory should significantly increase the number of comets discovered at distances beyond the range of solar observatories \citep[see e.g.][]{SilsbeeTremaine2016}, which will aid in future dedicated observing campaigns of comets. This, in turn, will make it easier to track comets' polarimetric signatures through time, and increase the chances of observing a comet at the same phase angle twice or more in polarimetric mode, especially in conjunction with observatories like \textit{STEREO-A}, i.e. far removed from the vantage point of the Earth.

\subsection{Phase Angle Curves}
\label{sec:phangle} 

\begin{figure}
\centering
\includegraphics[width=1.0\columnwidth,trim={1.6cm 8.8cm 1.6cm 1.1cm},clip]{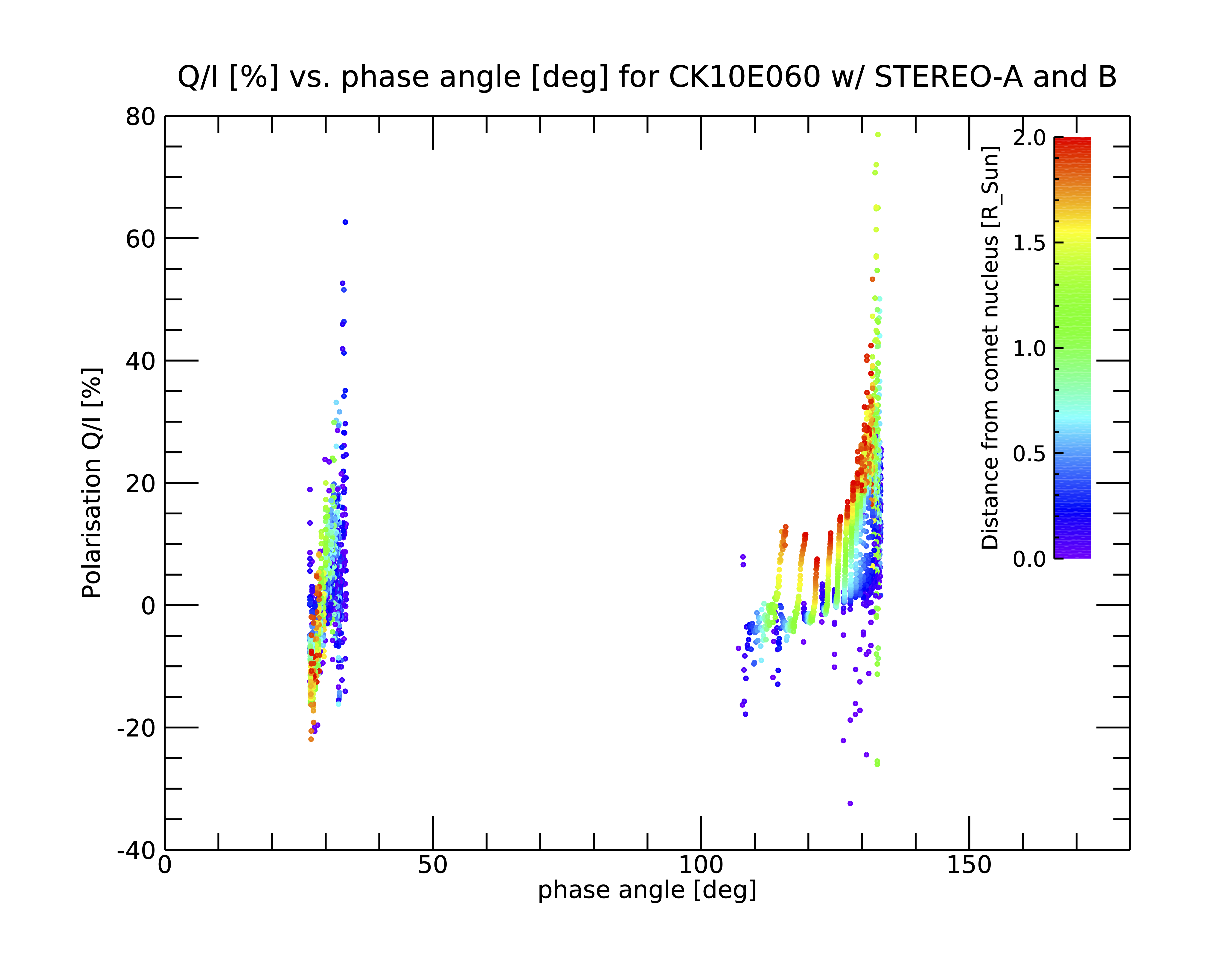}
\label{fig:iQU}
(i)

\centering
\includegraphics[width=1.0\columnwidth,trim={1cm 21cm 1cm 13cm},clip]{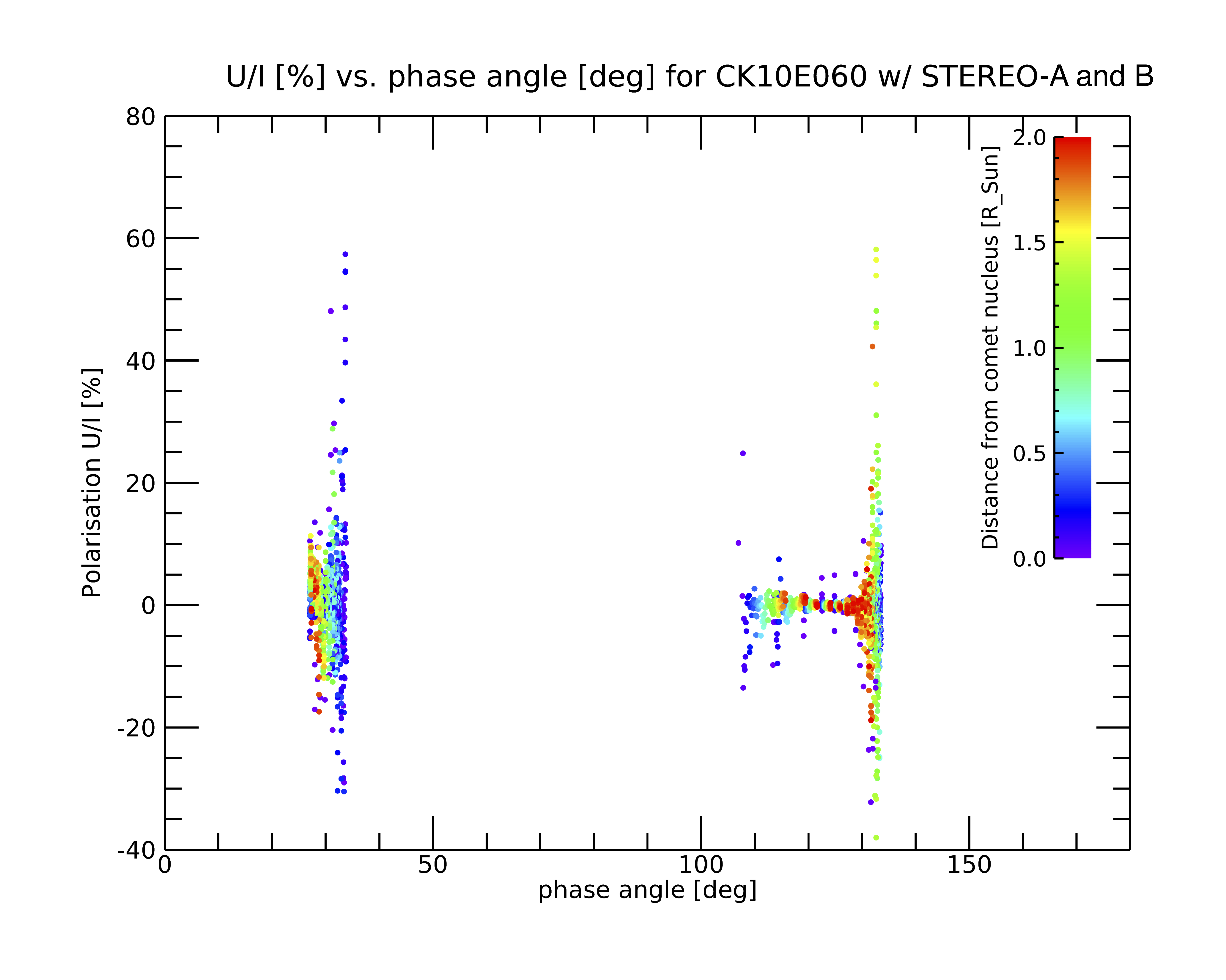}
\label{fig:iiQU}
(ii)

\caption{\small Plots of Stokes $Q/I$ (Fig. \ref{fig:CK10E060_ABQ}.i, above) and $U/I$ (Fig. \ref{fig:CK10E060_ABQ}.ii, below) vs. phase angle for all \textit{STEREO}/SECCHI/COR2 observations of comet C/2010 E6 with \textit{STEREO-A} (small phase angles) and \textit{B} (large phase angles) spacecraft. Each presented comet trace along the tail extends up to $2~R_{\odot}$ (red) from the comet photocentre (purple and dark blue, referred to as `comet nucleus' in the plot) via cyan, green, and yellow, depending on the quality of the imagery. The plotting procedure was inspired by \citet{Thompson2015}.}
\label{fig:CK10E060_ABQ}
\end{figure}

Figure \ref{fig:CK10E060_ABQ} shows $Q/I$  and $U/I$ for all the reduced comet tail imagery plotted against phase angle $\phi$ for the full set of \textit{STEREO-A} and \textit{STEREO-B}/SECCHI/COR2 observations of comet C/2010 E6. Each presented comet trace along the tail extends from the comet photocentre (purple and dark blue; changing colour along the way) up to $2~R_{\odot}$ (red) away from it, depending on the quality of the data. 

Focusing on the results themselves, \textit{STEREO-A} data is limited to a narrow range of phase angles -- between $\phi~{\sim}35^{\circ}$ and ${\sim}25^{\circ}$, decreasing with time while approaching perihelion (decreasing heliocentric distance). Despite the small range, the overall behaviour of the comet photocentre in this region is a clear change in polarisation from ${\sim}+5\%$ to ${\sim}-5\%$ in a steep curve with time, i.e. with decreasing phase angle and heliocentric distance. Generally, the comet tail increases in degree of polarisation with increasing distance from the comet. The quality of \textit{STEREO-A} data is lower than that of \textit{STEREO-B}. This can be fully attributed to differing geometry of observations. Figure \ref{fig:CK10E060_AB_orbits} shows the positions of both spacecraft and the orbit of comet C/2010 E6 (STEREO) in relation to the Sun, Earth, and other planets. It is clear -- particularly from the `side-on' view of the system -- that \textit{STEREO-B} is both generally closer to the comet and has the superior vantage point for observing the comet and its tail. From the vantage point of \textit{STEREO-A}, the tail appears highly projected, which hinders detailed analysis because we are receiving signal from grains at multiple distances in a single pixel. Due to this unfavourable geometry, the mapping of the data points onto the comet's orbital plane is also less accurate, contributing to lower phase angle resolution along the tail in each image and lower overall data quality. The data quality remains reasonably good when presented in other contexts (see Sec. \ref{sec:helio}). Fig. \ref{fig:CK10E060_ABQ}.ii shows a high degree of scatter of $U/I$ values in \textit{STEREO-A} data, especially at larger heliocentric distances (and larger phase angles) where the comet was dimmer.

Conversely, the quality of data derived from \textit{STEREO-B} observations of comet C/2010 E6 is high and only matched by a handful of other bright comets, e.g. C/2011 W3 (Lovejoy). The phase angle range observed is from $\phi~{\sim}135^{\circ}$ to ${\sim}105^{\circ}$, also decreasing with time while approaching perihelion. This $\phi$ range is higher than most comets have been observed at. The tail traces are clearly discernible in the figure as series of points tracing distinct curves (save a few stragglers) from dark blue at comet photocentre to red at $2~R_{\odot} $ and show a marked increase in degree of polarisation as we move away from the nucleus. The curve at the lowest end of the phase angle range shows unusual properties, but the overall trend at the inner coma is to change the polarisation from ${\sim}0\%$ to ${\sim}-5\%$ with time, i.e. decreasing phase angle and heliocentric distance. This is also unusual; the full meaning is considered in Sec.~\ref{sec:disc}. Fig. \ref{fig:CK10E060_ABQ}.ii shows a similar degree of scatter of $U/I$ at larger heliocentric distances (larger phase angles) in both \textit{STEREO-B} and \textit{A} data, both attributable to lower overall brightness of the comet and its tail. This trend is strongest for the tail data, with comet photocentre generally showing $U/I$ close to zero as expected. Overall, $U/I$ shows much less deviation from zero in \textit{STEREO-B} data when compared to \textit{A}, confirming the assertion that this data quality is higher. At smaller heliocentric distances (smaller phase angles) some scatter returns, mirroring the anomalous $Q/I$ data points seen there in Fig. \ref{fig:CK10E060_ABQ}.i.

While the observed phase angle $\phi$ changes somewhat with passage of time due to evolving observing geometries from both spacecraft, the variation in $\phi$ along the tail in each of the images is generally small. In \textit{STEREO-A} data, its variation along the tail is negligible. In \textit{STEREO-B} data, however, the variation changes from negligible at higher heliocentric distances (and higher $\phi$) to reaching ${\sim}10^{\circ}$ in the final images approaching perihelion (lower $\phi$).

\subsection{Photometry and polarisation vs. heliocentric distance}
\label{sec:helio} 
\begin{figure*}
\centering
\includegraphics[width=1.0\textwidth]{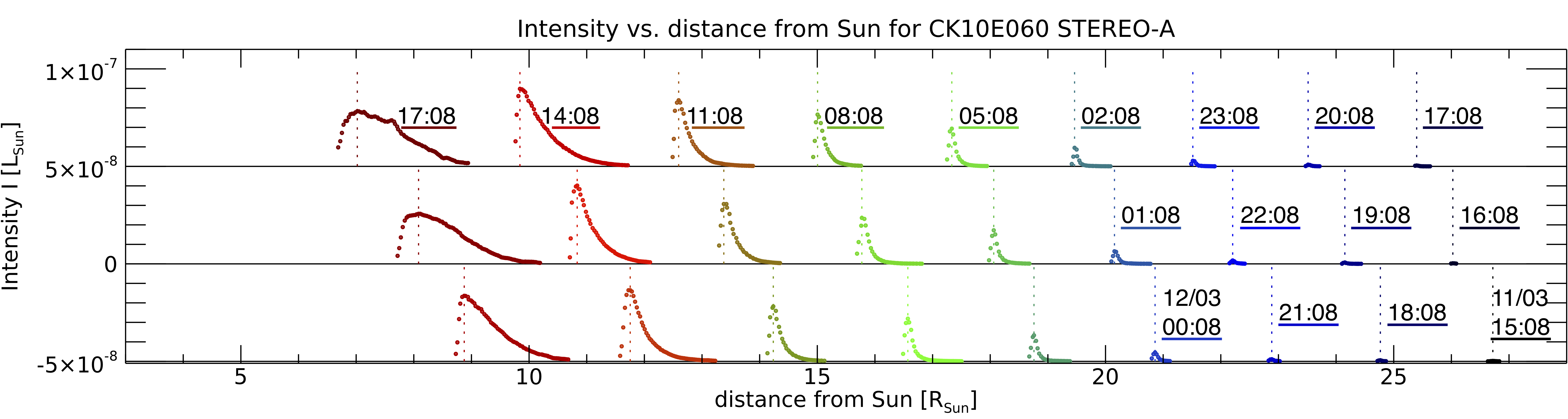}
\includegraphics[width=1.0\textwidth]{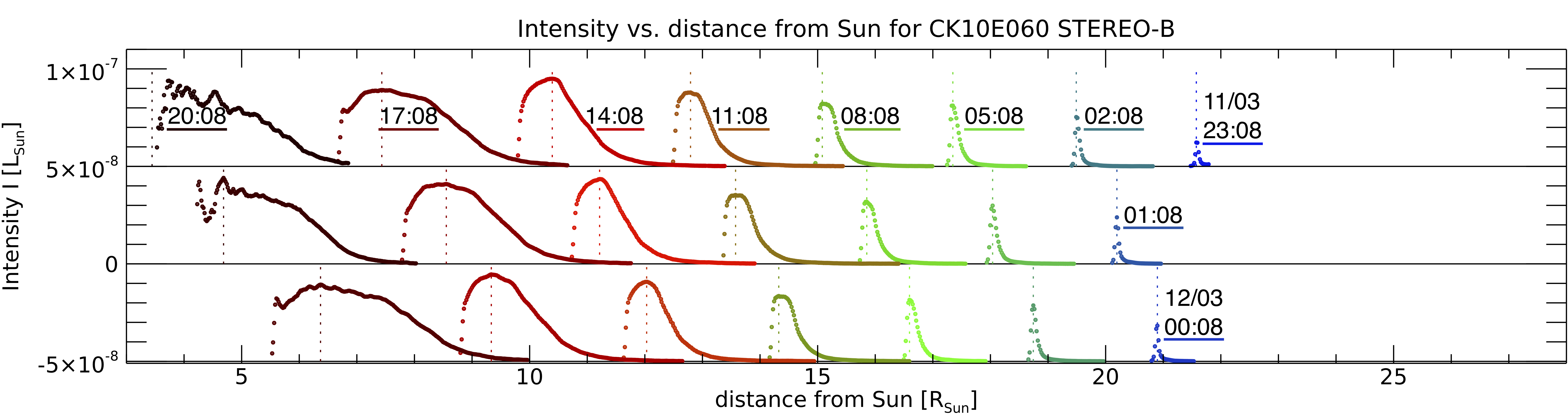}
(i)

\centering
\includegraphics[width=1.0\textwidth]{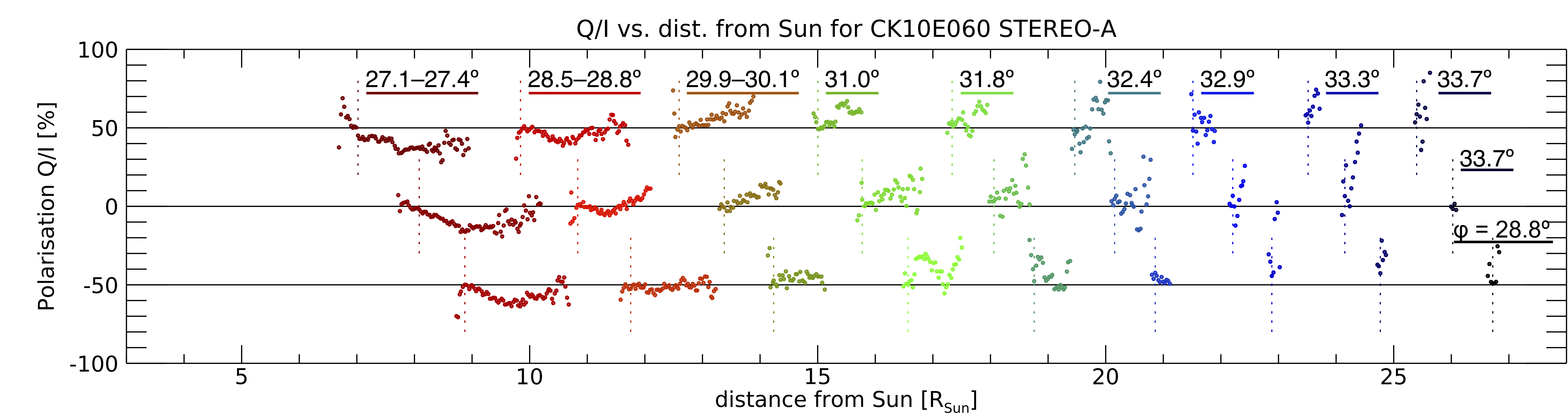}
\includegraphics[width=1.0\textwidth]{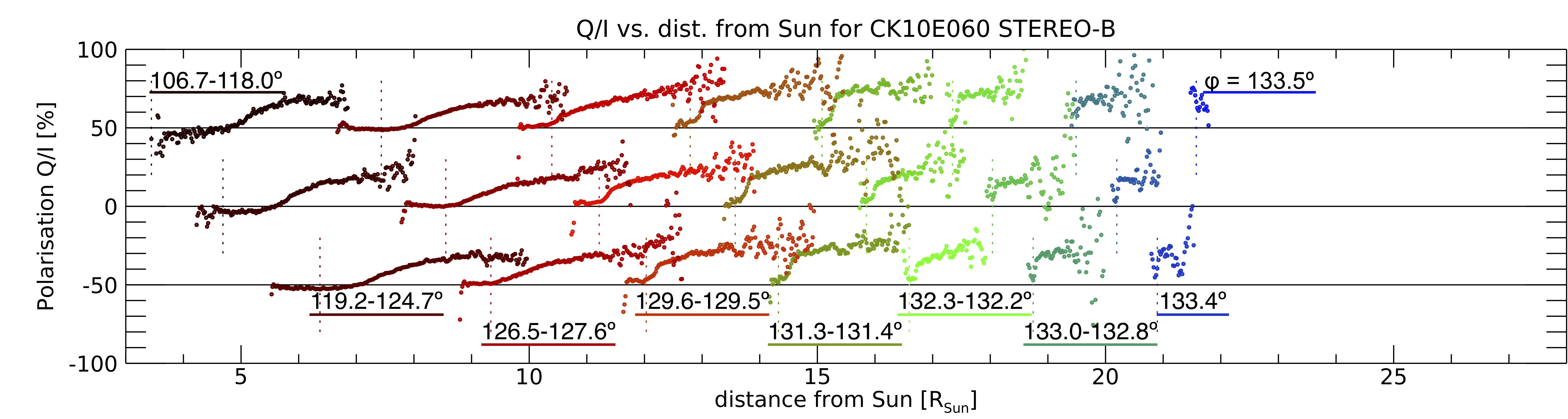}
(ii)

\caption{\small Plots of intensity (Stokes $I$, Fig. \ref{fig:CK10E060plotsIQI}.i, above) and $Q/I$ (Fig. \ref{fig:CK10E060plotsIQI}.ii, below) vs. heliocentric distance for the consecutive \textit{STEREO-A} (top) and \textit{B} (bottom) SECCHI/COR2 observations of comet C/2010 E6 (STEREO) and its tail, starting on the right-hand end (11th March 2010 at 15:08 UT for \textit{STEREO-A}, at 23:08 UT for \textit{STEREO-B}, as annotated in Fig. \ref{fig:CK10E060plotsIQI}.i) and moving closer to the Sun in hourly increments. For clarity of presentation, the data is sequentially offset by $-4\cdot10^{-8}~L_{\odot}$, zero, $4\cdot10^{-8}~L_{\odot}$, etc. in Fig. \ref{fig:CK10E060plotsIQI}.i or $-50~\%$, $0~\%$, $+50~\%$, etc. in Fig. \ref{fig:CK10E060plotsIQI}.ii, creating three levels. The colour gradient, related to heliocentric distance (dark blue at large ones, dark red closer to the Sun), is intended as an aid to visualising that the data -- even though presented at three separate levels -- is a continuous set. The solid black lines indicate zero polarisation for each of the offsets. The coloured vertical dashed lines indicate the position of maximum intensity along the comet tail for each respective observation. The phase angle $\phi$ at the comet photocentre, or a range from photocentre to tail, is presented for selected observations in Fig. \ref{fig:CK10E060plotsIQI}.ii.}
\label{fig:CK10E060plotsIQI}
\end{figure*}

Figure \ref{fig:CK10E060plotsIQI} shows the full set of Stokes $I$ and reduced $Q$ data on comet C/2010 E6 (STEREO) with respect to heliocentric distance, analysed from the \textit{STEREO}/SECCHI/COR2 imagery. They start on the right-hand end (11th March 2010 at 15:08 UT for \textit{STEREO-A}, at 23:08 UT for \textit{STEREO-B}) and move closer to the Sun in hourly increments. Since the data would contain considerable overlap otherwise, three different offsets are used, along with a colour gradient (dark blue at large heliocentric distances, dark red at small ones) to further emphasise the fact that plots, even if offset, are part of the same series. 

For example -- and one can consult either Figure \ref{fig:CK10E060plotsIQI}.i or \ref{fig:CK10E060plotsIQI}.ii here with the same effect -- \textit{STEREO-A} observations begin on 11th March 2010 at 15:08 UT at heliocentric distance of ${\sim}26.8~R_{\odot}$ (rightmost dataset, bottom level, black/very dark blue, peak intensity denoted by the dashed vertical line of the same colour). The next dataset, at heliocentric distance of ${\sim}26.1~R_{\odot}$, at the middle level (also black/very dark blue), was taken at 16:08 UT, etc. At 23:08 UT (third dataset from the right on the top level in \textit{STEREO-A} data, at  ${\sim}21.6~R_{\odot}$, blue), simultaneous observations begin, and the \textit{STEREO-B} observations are also at  ${\sim}21.6~R_{\odot}$, on the top level, and blue. Simultaneous observations from both spacecraft end on 12th March 2012 at 17:08 UT (peak intensity at  ${\sim}7~R_{\odot}$, top level, dark red). Reliable \textit{STEREO-B} observations continue for three more hours, until 20:08 UT and peak intensity distance of  ${\sim}3.4~R_{\odot}$ (black/very dark red). Each set of points of the same colour belongs to observations taken at the same time. While phase angle information is lost in this plot, its variability along the comet tail is almost negligible and trends with changing cometocentric distance are clear; consult Sec. \ref{sec:phangle} for details.

Focusing specifically on Figure \ref{fig:CK10E060plotsIQI}.ii, we can better appreciate the steep slope of tail polarisation as we move away from the comet coma, especially in the early observations (right-hand end of \textit{STEREO-A} data, dark blue). Increase of polarisation with cometocentric distance is a known phenomenon \citep[e.g.][]{PolarComets}, but such a steep slope has only been observed for other near-Sun comets \citep[][Figs. 9 and 7, respectively]{Thompson2015,Thompson2019paper}. Both the reasons for this trend and the deviations from it will be discussed in the next section.

\section{Discussion}
\label{sec:disc} 

\subsection{Comparison to ground-based comet polarimetry}

\begin{figure}
\centering
\includegraphics[width=1.0\columnwidth]{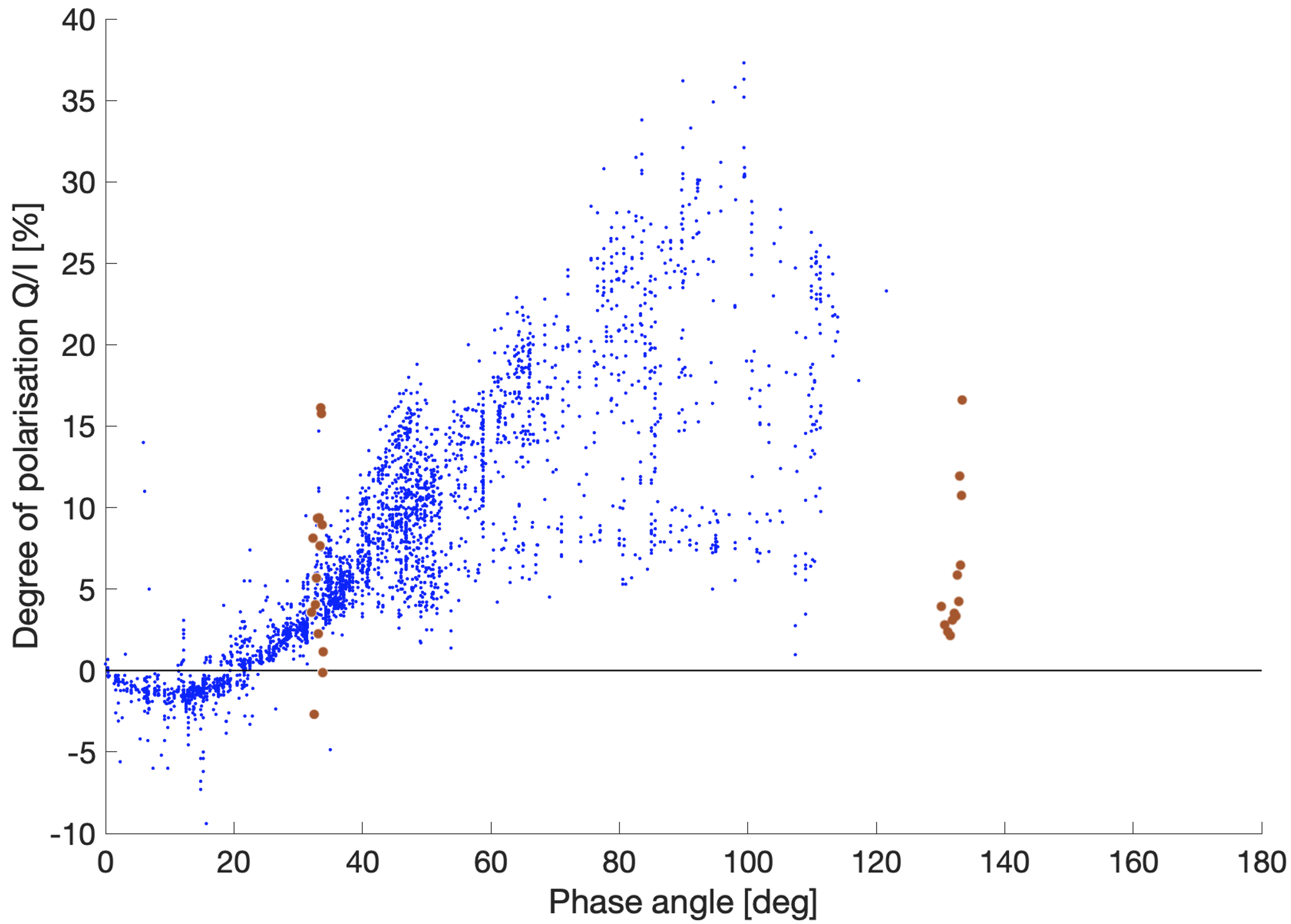}
\caption{\small The full dataset for polarisation of comet comae from the newest Comet Polarimetry Database \citep{Kiselev2017compilation} presented as a polarimetric phase angle plot (small blue dots). A variety of telescopes, filters, and analysis techniques has been used. From the dataset, only comet Ikeya-Seki has been excluded, since its observations comprised the tail rather than the coma \citep{WeinbergBeeson1976,BappuEtAl1967}. Averaged comet photocentre polarisation of comet C/2010 E6 (STEREO) from this work has been added for contrast (larger filled brown circles).}
\label{fig:AllCometsCPDKiselev}
\end{figure}

Both STEREO-A and B observations show some peculiar behaviour, which must first be compared against the wealth of existing data from generally ground-based comet polarisation observations. Figure \ref{fig:AllCometsCPDKiselev} shows the full dataset from the newest Comet Polarimetry Database \citep{Kiselev2017compilation} in a polarimetric phase angle plot. Plotted are the data points of comet coma polarisation vs. phase angle for 73 different comets with over 3000 measurements spanning over a century between them, and yet a clear trend can be seen. 

The comet comae show negative polarisation between phase angles $\phi = 0^{\circ}$ and $22^{\circ}$ (called the inversion angle) -- this regions is referred to as the negative polarisation branch -- with a minimum near $Q/I=-1.5~\%$. Polarisation appears to reach a peak around $\phi = 90 - 100^{\circ}$, after which it drops again, presumably smoothly returning to zero at $\phi = 180^{\circ}$. There are essentially no data points beyond $\phi = 120^{\circ}$, however, due to observing limitations discussed earlier. A handful of comets have been observed and analysed at higher phase angles with \textit{SOHO} and \textit{STEREO} spacecraft but have not been recorded in the Comet Polarimetry Database. \cite{Hui2013} analysed \textit{STEREO-B} observations of comet P/2003 T12 = 2012 A3 (SOHO) with phase angles exceeding $\phi = 170^{\circ}$ and thus cannot be directly compared to results presented here. \cite{GrynkoEtAl2004} analysed \textit{SOHO} observations of comet 96P/Machholz in a broader range of phase angles and found generally good agreement with high polarisation data in Fig. \ref{fig:AllCometsCPDKiselev}, although with a large degree of uncertainty related to the difficulties with \textit{SOHO} data discussed in Sec. \ref{sec:obs}. Notably 96P/Machholz is a periodic comet, therefore more highly processed than Kreutz comets, and -- with perihelion at ${\sim}0.124$~AU -- merely a sunskirter rather than a sungrazer \citep{GeraintEtAl2018}.

The degree of polarisation at the peak varies significantly; from $5~\%$ to near $40~\%$. This variation likely reflects the composition of the comets; higher maximum of polarisation is found for dust-rich comets, lower for gas-rich comets. This is easily explained, as light scattering from gases will generally suppress the polarisation induced by the dust particles \citep{PolarComets} - see \ref{sec:nearSun} for how this contrasts with observations of near-Sun comets. Similar polarimetric properties are found for asteroids (particularly C-type) and zodiacal dust, hinting at a common origin of the dusty material \citep{LevasseurRegourdEtAl1990}. 

Comparing results from the Comet Polarimetry Database against the phase angle curves of comet C/2010 E6 (STEREO), there are clear departures from the trends presented here. Focusing on the coma region (as presented in Fig. \ref{fig:AllCometsCPDKiselev}) for the most direct comparison, \textit{STEREO-A} observations fall within the phase angle range of previous observations and do show both negative polarisation and an increase of polarisation with increasing $\phi$, however the inversion angle is at $32^{\circ}$ rather than $22^{\circ}$, and the slope is much higher than expected. \textit{STEREO-B} observations mostly sit beyond the usual range of phase angles, but they also show an increase with increasing phase angle, where a decrease from the peak polarisation at $\phi = 90 - 100^{\circ}$ is expected. The solution for these discrepancies is that both of these sets of observations are greatly affected by the near-Sun environment: for \textit{STEREO-A} observations, the heliocentric distance varies greatly through the observing run, while the phase angle change remains minimal due to geometry of observations (see Fig. \ref{fig:CK10E060_AB_orbits}). For \textit{STEREO-B} observations, the change in heliocentric distance induces an overall decrease in polarisation with time, but this happens to correlate with decreasing phase angle. Both of these effects will be further discussed in Section \ref{sec:nearSun}. A caveat in this comparison is that the methodology of observations in \cite{Kiselev2017compilation} is different from our own, however our observations show behaviour extreme enough that it is difficult to compare to any broad review of polarimetric properties of comets.

\subsection{Effects of the near-Sun environment}
\label{sec:nearSun} 

There are a number of effects at play at the small heliocentric distances where comet C/2010 E6 (STEREO) was located when observed by the twin \textit{STEREO} spacecraft. The comet nucleus will be affected by sublimation of refractory material, which may cause ablation of the surface or rotational spin-up and break-up of the object due to insufficient tensile strength. Tidal forces may play a role. The effect of the diamagentic cavity in shielding the comet from the effects of solar wind -- diminished as the comet approaches the Sun -- should also be considered. The thermal wave may propagate inside the nucleus -- especially if the mantle of old ejected material is not being replenished due to sublimation effects -- and disintegrate it via sublimation of unexposed ices and refractory material \citep[and references therein]{GeraintEtAl2018}. Realistically, all of these and other effects are likely to play some role in the evolution and final fate of the comet in the near-Sun environment.

Assuming, firstly, a radiative cooling model based on black body radiation for the comet nucleus, the local temperature can be compared to the equilibrium temperature of various materials. For example, the equilibrium temperature for carbon dioxide is reached at a heliocentric distance of $10$~AU, and for water at just under $3$~AU ($273.16$~K). The proximity to the Sun can promote phase changes of refractory material and cometary dust itself in addition to the ices. This will affect its polarimetric properties. Refractory organics begin decomposing and sublimating at ${\sim}450$~K, reached around 0.7~AU (${\sim}140~R_{\odot}$), and silicates in a range between $1000-1500$~K (distance between ${\sim}0.07-0.047$~AU or ${\sim}14-10~R_{\odot}$), depending on their composition. Forsterite, for example (in the olivine family), is expected to start sublimating at heliocentric distances below ${\sim}0.015$~AU; $3~R_{\odot}$ \citep{KimuraEtAl2002,GeraintEtAl2018}.

Recalling that the comet tail tends to show an increase in polarisation with increasing distance from the comet nucleus, this is a fact previously observed for other comets \citep[e.g.][]{PolarComets}, though it can be modulated by decreases in polarisation due to, presumably, variation in the material ejected from the nucleus. This was seen in some polarimetric observations of comet 67P/Churyumov-Gerasimenko \citep[e.g.][]{HadamcikEtAl2016,RosenbushEtAl2017,Nezicthesis2020}. The slope, however, is much higher for comet C/2010 E6 (STEREO) than for comets at larger heliocentric distances. Similar result was found by \citet{Thompson2015} for comet C/2011 W3 (Lovejoy). 

The standard explanation for increasing polarisation with cometocentric distance is that dust particles are processed over time as they move away from the nucleus. Assuming a fluffy aggregate composition with silicate grains and refractive organic matrix \citep{KimuraEtAl2002,KimuraEtAl2006,Kolokolova2016}, this will, in general, result in smaller grains with various materials slowly removed by a variety of processes.

Since the organic matrix is the first refractory material to begin sublimating, it is reasonable to assume that coronagraph observations of comet C/2010 E6 (with heliocentric distances $\ll 140~R_{\odot}$) already see the tail mostly depleted of it, except in the newly ejected material near the comet photocentre. This causes a particularly steep gradient in polarisation, as silicates become the dominant species as the organics are depleted, in addition to the normal processing of the material with time creating smaller particles. A similar hypothesis is put forward by \cite{Thompson2019paper}.

In addition to this, presence of amorphous carbon among the organics may have a significant effect, as its sublimation temperature is very high. Amorphous carbon is created by UV irradiation of the organics, and is much more strongly affected by the solar radiation pressure than the silicate particles \citep{ZubkoEtAl2015}. This means amorphous carbon is likely to be swept along the tail. Theoretical modelling by \cite{ZubkoEtAl2013} shows that such material can show very strong polarisation. This could easily explain the observed increase in polarisation along the tail, especially the steep slope at higher heliocentric distances.

Analysis of Fig. \ref{fig:CK10E060plotsIQI} shows us a broader picture of the comet's behaviour over time. While the tail (in \textit{STEREO-B} data in particular) still increases in polarisation over time far from the nucleus, a disruption is propagating from the coma and along the tail. This is seen in intensity plots as the broadening of the intensity peak (\textit{STEREO-B}) or, less clearly, as the decrease in the intensity drop-off slope (\textit{STEREO-A}). The overall peak brightness of the comet, $4\cdot 10^{-8} L_{\odot}$, is reached between $9$ and $12~R_{\odot}$, which agrees with the analysis of Kreutz group comets in \textit{SOHO} coronagraph imagery \citep{BieseckerEtAl2002,KnightEtAl2010}. \citet{KimuraEtAl2002} also found that fluffy aggregates composed of olivine are likely to cause a peak in brightness between $11.2-12.3~R_{\odot}$, which falls within this region, with sublimation of olivines at smaller heliocentric distances being the dominant process decreasing coma brightness. An additional brightness enhancement and eventual decrease within $7~R_{\odot}$ is attributed to the presence and then sublimation of pyroxenes, but those distances are sampled less well in this analysis.

This disruption is also characterised by a clear decrease in polarisation to near-zero for \textit{STEREO-B} data, and in appearance of negative polarisation for \textit{STEREO-A} -- for both the coma and the tail. 

The evidence for this disruption is circumstantial, since the resolution of the instruments is insufficient to discern variation in morphology of the comet coma, but similar photometric (broadening of the brightness peak) and polarimetric (an abrupt decrease in polarisation near the coma vs. the tail) effects have been observed in other Kreutz sungrazers, e.g. comet C/2011 W3 (Lovejoy) \citep{Thompson2015,Nezicthesis2020}, which is known to have experienced extreme disruption at its perihelion approach \citep[e.g.]{SekaninaChodas2012}. We also know that comet C/2010 E6 (STEREO) did not reappear after perihelion, as is the case for most Kreutz comets due to their small size. A fragmentation of the nucleus, combined with rapid sublimation of newly exposed material, is therefore the most likely cause of the broadening of the intensity peaks at decreasing heliocentric distances.

In light of the fragmentation hypothesis, the negative polarisation seen in \textit{STEREO-A} data may be caused by the combination of new, freshly exposed silicate particles ejected from the nucleus on the one hand, and the low phase angle on the other. While the inversion angle is expected at $22^{\circ}$ and is instead seen at $32^{\circ}$, a small variation in the structure and composition of the particles can extend the theoretical models of the negative branch even to near $40^{\circ}$ \citep[e.g.][]{PetrovaEtAl2001,FrattinEtAl2019,ZubkoEtAl2020,HalderGanesh2021}. Some of the aforementioned theoretical models reproduce the results more readily with compact particles rather than fluffy aggregates, although \citet{ZubkoEtAl2015JQSRT} argues that packing density does not have a significant effect on the polarimetric response when the particle morphology attains a significant level of disorder, when compared to the effects of refractive index. A strong negative polarimetric response at a variety of phase angles might be caused, for instance, by particles with rounded shapes caused due to melting \citep[e.g.][]{HansenTravis1974,HansenHovenier1974}. 

The abrupt reduction of polarisation to small positive polarisation in \textit{STEREO-B} data near side-scattering, however, is not quite as easily explained. The often-used explanation of surrounding gas causing the dampening of the polarimetric response is unlikely to be valid for this feature due to the close proximity to the Sun, where strong dissociation effects mean most gas species are very short-lived \citep[e.g.][]{BieseckerEtAl2002,Luk'yanykEtAl2020}. \citet{BieseckerEtAl2002} has shown that some species, such as NaI, may be detected as close as $7~R_{\odot}$, with others, like Ly$\alpha$, appearing even closer to the Sun, but neither of those would be detected by the COR2 coronagraph. Comets observed near $1$~AU have several common sources of emission in the COR2 bandpass, including C2, NH2, and forbidden O \citep[e.g.][]{FeldmanEtAl2004} but, as noted above, all are expected to be short-lived
at the heliocentric distances of our observations. Some theoretical light-scattering models can produce small positive polarisation at phase angles seen in \textit{STEREO-B} data while also showing negative polarisation in the backscattering regime \citep[e.g.][]{ZubkoEtAl2014,FrattinEtAl2019,HalderGanesh2021}. Those results do not match the observed behaviour perfectly, but they do approach it closely, and most of them match the results best with Mg-rich silicate particles such as forsterite. This reconciles well with the photometric observations and is therefore the most likely resolution, although future theoretical modelling that reproduces the observations even better would be beneficial. A role may even be played by the peculiar observing conditions, especially the spread of phase angles due to the Sun acting as an extended source -- see Sec. \ref{sec:extendedSun} above -- though the effects of this are not entirely clear.

This difference in polarimetric behaviour in simultaneous observations of a sungrazing comet can also be observed in post-perihelion observations of comet C/2011 W3 (Lovejoy): \textit{STEREO-B} (phase angle range $115-120^{\circ}$) observed near-zero polarisation of the re-emerging comet tail, while \textit{STEREO-A} (phase angle range $32-34^{\circ}$) observed negative polarisation in the older parts of the tail (i.e. the ones originating closer to the Sun) \citep{Nezicthesis2020}. A similar solution should apply to those observations.

A close look at \textit{STEREO-B} data in Fig. \ref{fig:CK10E060plotsIQI}.ii shows that near-zero polarisation in the coma region of comet C/2010 E6 (STEREO) might be traced back to ${\sim}20~R_{\odot}$ (best seen in \textit{STEREO-B} data), which precedes the change in the shape and peak of the intensity curve by several hours and solar radii. This may simply be an artefact of data analysis, as small variations in mutual alignment of sharp polarised intensity peaks can have significant effects on polarimetric calculations. It is difficult to decouple that from the emerging broad intensity peak which also reduces polarisation to zero. Other studies have found similar effects, however; \cite{Sekanina2000b} found that dust production peaked at $20-30~R_{\odot}$ in a sample of 9 comets, and \cite{KnightEtAl2010} similarly sees a dramatic change in slope of comet brightening around the same heliocentric distance. This effect may indicate a thus far unexplored process for suppressing the polarimetric response: the heliocentric distance is likely too high to indicate sublimation of silicates at that point. It might, however, be caused by fragmentation of the nucleus itself. This would expose the water ice within and likely suppress the polarisation signal. While this is occurring much earlier than would expected due to tidal forces \cite{KnightWalsh2013}, there appears to be some precedent to that, as comet C/2012 S1 (ISON) has shown indications of fragmentation at heliocentric distances as high as $0.6$~AU or $129~R_{\odot}$ \citep{SekaninaKracht2014,BoehnhardtEtAl2013}. It is therefore plausible that these near-Sun comets are more susceptible to fragmentation than previously thought due to their internal structure; perhaps the thermal wave propagates faster or deeper than for e.g. Jupiter-family comets. It is also plausible that this event caused the propagation of the broad intensity peak and low polarisation, indicating further break-up of the comet.

\section{Conclusions}
\label{sec:concl} 

Photometric and polarimetric observations of the sungrazing Kreutz comet C/2010 E6 (STEREO) by the twin \textit{STEREO} spacecraft in March 2010 were analysed and the implication of the results discussed. Since the comet was not observed post-perihelion, and since this is the fate of most Kreutz comets, it can be safely assumed that the comet fully disintegrated during its perihelion passage. Prior to that, however, it exhibited a variety of peculiar behaviours -- both photometric and polarimetric --  which distinguish it from most other comet observations (usually at higher heliocentric distances).

Effects of non-negligible angular size of the Sun at small heliocentric distances under consideration were carefully scrutinised. The values reported in this work will be dominated by signal near the mean values, reducing any potential observable effects. Variation in scattering plane angle was found to cancel out effects on $U/I$ by symmetry, whereas the effects on $Q/I$ were found to be very small. The uncertainty of the phase angle has also not shown any significant effects on the results, though such effects are more difficult to decouple from the overall peculiar behaviour of the comet. Analysis of a near-Sun comet with more conventional polarimetric characteristics such as 96P/Machholz, which is beyond the scope of this investigation, may shed more light on the potential effects the increased angular size of the Sun might have. For 96P/Machholz in particular the angular size of the Sun at perihelion is only ${\sim}4^{\circ}$, however, which may not be enough for a discernible result. We recommend observations of a non-disintegrating comet at multiple heliocentric distances but similar phase angles as the best candidate for resolving these effects; a technically challenging task.

Most notable of the peculiar behaviours of comet C/2010 E6 (STEREO) is the sudden drop in polarisation of the comet nucleus observed in \textit{STEREO-B} data (potentially from heliocentric distances of ${\sim}20~R_{\odot}$), which then spread gradually along the comet tail. This was accompanied (below ${\sim}15~R_{\odot}$) by a broadening of the the intensity peak from the nucleus further down the tail. The broadening and brightening is consistent with that observed by e.g. \citet{KnightEtAl2010} for a typical Kreutz comet and is likely a result of fragmentation of the nucleus, but the drop in polarisation gives us an earlier indication of the changing processes in the near-nucleus region, proving it is a useful tool for analysis in this extreme environment. 

The observed polarimetric signature of near-zero positive polarisation (\textit{STEREO-B}, high phase angles) and, simultaneously, negative polarisation (\textit{STEREO-A}, small phase angles) spreading from the near-nucleus region along the tail may be best explained by the presence of Mg-rich silicate particles, the amount of which has been increased due to the fragmentation of the nucleus, and which are expected to be the prevailing component at such small heliocentric distances and equilibrium temperatures. Although the theoretical light scattering models do not fully match the observed results, this explanation best accounts for the observed polarimetric and the photometric behaviour of comet C/2010 E6 (STEREO).

The observations at varying distances clearly show us the changing behaviour of the comet and thus help shed more light on its likely structure and composition. The benefit of near-simultaneous polarimetric observations from different phase angles is also clear, as any hypotheses are required to explain both sets of observations at once.

It is reasonable to assume that comet C/2010 E6 (STEREO) is, in its peculiar behaviour, a typical representative of the Kreutz group. Further analysis of similar observations of near-Sun comets -- already commenced on a set of seven additional comets within \citet{Nezicthesis2020} -- and theoretical modelling of light-scattering properties of cometary dust particles will be required to shed more light on this topic. Recent advances -- like the work of \cite{HalderGanesh2021} -- show great promise on the latter point, building upon the work of past decades. Thousands of comets have been observed by SOHO \cite{BattamsKnight2016} and hundreds by STEREO spacecraft, and although majority of them are too faint for a thorough polarimetric analysis, it is likely that several dozen of them are bright enough to analyse and combine into a more coherent picture of the Kreutz family population and effects of the near-Sun environment.

\section*{Acknowledgements}
This research has made use of data and/or services provided by the International Astronomical Union's Minor Planet Center. It would not have been possible without the publicly available \textit{STEREO} data, or the work done by the Sungrazer Project. It was supported by a grant from the UK Science and Technology Facilities Council. The authors would also like to thank the reviewer for their valuable insights, and Mark Bailey for fruitful discussions. 

\section*{Data Availability}
The raw observational data is freely available from e.g. \url{https://stereo.nascom.nasa.gov/data/}. The details of the analysis procedure and the processed data underlying this article will be shared on reasonable request to the corresponding author.



\bibliographystyle{mnras}
\bibliography{0_PAPER}








\bsp	
\label{lastpage}
\end{document}